\documentclass[%
 reprint,
 amsmath,amssymb,
 aps,
]{revtex4-2}

\usepackage{graphicx}
\usepackage{dcolumn}
\usepackage{bm}
\usepackage{hyperref} 
\usepackage{color}
\usepackage{soul}
\usepackage{float}
\usepackage[dvipsnames]{xcolor}
\usepackage{physics}
\usepackage{ulem}
\graphicspath{{./}}
\usepackage{unicode-math}
\usepackage{makecell,tabularx,multirow}
\usepackage{upgreek}

\usepackage{romannum}

\begin{document}

\preprint{APS/123-QED}

\title{Nonlinear optical charge state switching and pumping to a diamond NV center dark state}
\author{Prasoon K.\ Shandilya}
\thanks{These authors contributed equally.}
\author{Vinaya K.\ Kavatamane}
\thanks{These authors contributed equally.}
\author{Sigurd Fl{\aa}gan}
\thanks{These authors contributed equally.}
\author{David P.\ Lake}
\author{Denis Sukachev}
\author{Paul E.\ Barclay}

\email[Paul~E.\ Barclay: ]{Corresponding author pbarclay@ucalgary.ca}

\affiliation{Institute for Quantum Science and Technology, University of Calgary, Calgary, Alberta T2N 1N4, Canada}

\date{\today}

\begin{abstract}
The photodynamics of diamond nitrogen-vacancy (NV) centers limits their performance in many quantum technologies. Quenching of photoluminescence, which degrades NV readout, is commonly ascribed to a dark state that is not fully understood. Using a nanoscale cavity to generate intense infrared fields that quench NV emission nonlinearly with field intensity, we show that the dark state is accessed by two-photon pumping into the $^4\!A_2$ quartet state of the neutrally charged NV (NV$^0$).
We constrain this state's energy relative to the NV$^0$ ground-state ($^2\!E$) to ${<}0.58$\,eV and the recombination energy threshold to the NV$^-$ ground state ($^3\!A_2$) to $\leq2.33\,\text{eV}$.
Furthermore, we estimate the intrinsic lifetime of $^4\!A_2$ state to be $1.78-6.06\,\upmu\text{s}$ and show that accessing this state allows sensing of local infrared fields. This new understanding will allow predictions of the limits of NV technologies reliant upon intense fields, including levitated systems, spin--optomechanical devices, and absorption--based magnetometers.
\end{abstract}

\maketitle
Diamond nitrogen-vacancy (NV) color centers are a promising platform for quantum technologies\,\cite{Awschalom2018,Ruf2021JAP, Barry2020}. Their optically addressable electronic spins can be used for quantum memories\,\cite{Dutt2007,Fuchs2011} and distributed quantum entanglement\,\cite{Hensen2015,Pompili2021, Hermans2022}, and they have enabled imaging of biological\,\cite{Schirhagl2014,Zhang2021Sensors} and condensed matter\,\cite{Casola2018,Hedrich2021} systems via electric\,\cite{Dolde2014PRL}, magnetic\,\cite{Maze2008,Rondin2014}, and temperature sensing\,\cite{Neumann2013}. Hybrid devices formed by integrating NV centers with nanomechanical resonators\,\cite{Teissier2014, Macquarrie2013,Lee2017, Shandilya2021, Shandilya2022} and superconducting circuits\,\cite{Kubo2010} are leading to quantum technologies for transduction and on-chip qubit networking\,\cite{Marcos2010, Neuman2021}. To further advance these applications, it is essential to understand the physical mechanisms affecting the NV center's optical and electronic properties.

The NV center is commonly observed in two charge states: negatively charged NV$^{-}$ and neutrally charged NV$^{0}$ \cite{Doherty2013}. The deeply studied spin and optical properties of NV$^{-}$ are central to many quantum technologies, and NV$^0$ has enabled charge-based memories\,\cite{Dhomkar2016,Jayakumar2016} and enhanced readout of NV$^-$ spins\,\cite{Shields2015,Jaskula2019}. While NV$^0$ and NV$^-$ have well characterized photoluminescence (PL) spectra, a poorly understood non-fluorescent state, referred to as a `dark state', is observed in some excitation schemes\,\cite{Han2010,Waldherr2011,Beha2012,Geiselmann2013}. Insight into charge states can be obtained through multi-wavelength excitation, including photons with energy smaller than NV optical transitions but sufficient for charge conversion. For example, combining near-infrared (IR) and IR (721--1064\,nm, 1.720--1.165\,eV) with green (532\,nm, 2.330\,eV) excitation modifies the charge state dynamics and PL of NV centers\,\cite{Lai2013, Neukirch2013, Hopper2016, Ji2016, Hacquebard2018, Roberts2019, RamanNair2020, Qian2022PRA}. A striking application of charge state manipulation is quenching of NV center emission by near-IR fields in stimulated emission depletion (STED) super-resolution microscopy\,\cite{Rittweger2009}. Despite progress in understanding NV charge-state dynamics, a model describing quenching and the corresponding dark state is elusive. 

\begin{figure}[t!]
	\includegraphics[width=\linewidth]{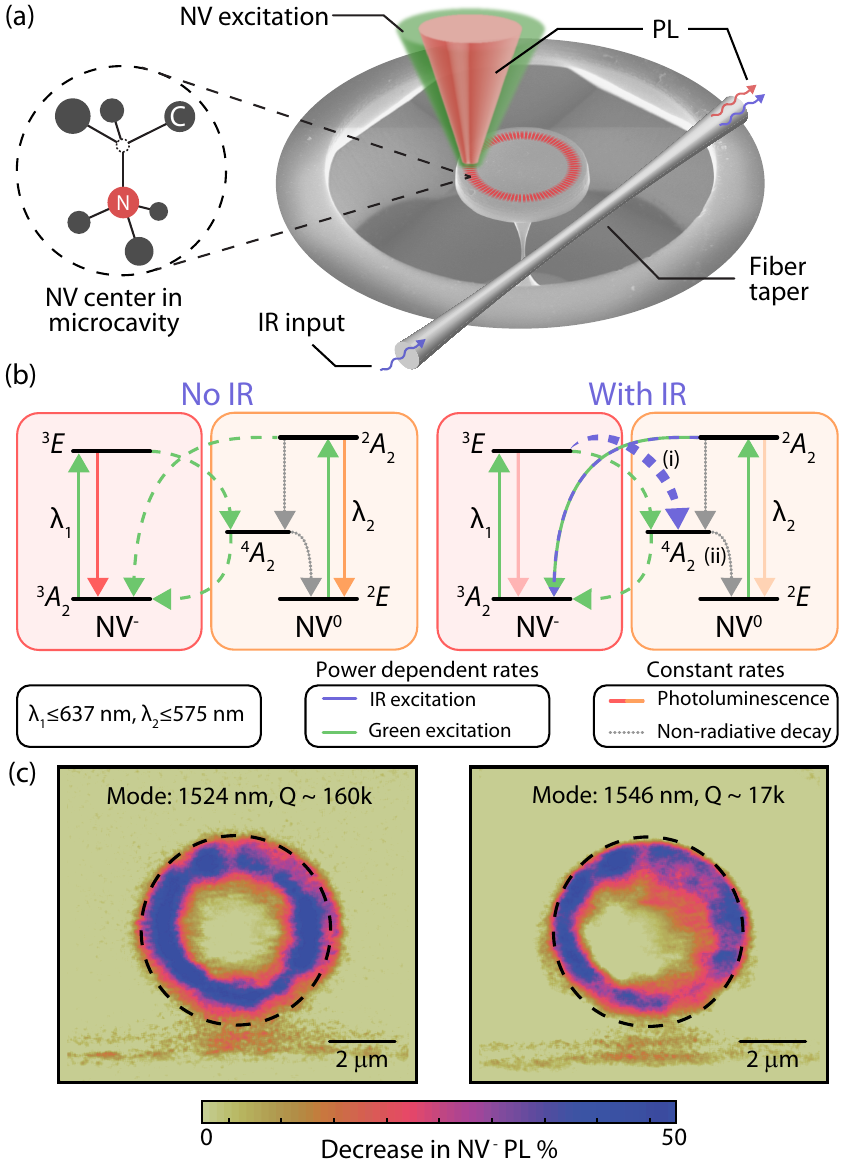}
		\caption{
		    \label{fig1}
		    Combined microcavity (IR) and confocal microscope (532\,nm) excitation of NV centers. 
            (a) Scanning electron micrograph of a microdisk overlayed with a measurement schematic. A confocal microscope excites NV centers with green light. PL is collected by the microscope or the fiber taper evanescently coupled to the microdisk. The fiber taper also couples IR light into the device.
            (b) Model of the charge-state dynamics under green (left) and simultaneous green and IR (right) excitation.
            Two IR photons (i) bring NVs pumped by the green laser from the negatively charged $^3E$ excited state to the neutrally charged $^4\!A_2$ state. Non-radiative decay and green excitation (ii) relaxes NVs to the NV$^0$ and NV$^-$ orbital ground-states, respectively. At high IR power, the nonlinear optical rate (i) to $^4\!A_2$ is faster than non-radiative decay (ii) to $^2\!E$, trapping population in this dark state. Photoionization requires excitation of NV$^-$ ($\lambda_{1}\leq637\,\text{nm}$ (1.946\,eV)); recombination requires excitation of NV$^0$ ($\lambda_{2}\leq575\,\text{nm}$  (2.156\,eV)). 
            Dashed lines indicate transitions between charge states.
            (c) Spatially resolved change in NV PL induced by IR fields in the microdisk. 
		    }
\end{figure}

In this Letter, we show that the NV dark state can be populated through nonlinear optical excitation by IR fields. Using diamond microdisks supporting high quality factor ($Q$) whispering gallery modes (Fig.\,\ref{fig1}\,(a)), we study NV center PL quenching for IR field strengths varying over six orders of magnitude at two wavelengths (966\,nm and 1524\,nm). Comparing measurements with a model by Razinkovas et al.\,\cite{razinkovas2021photoionization} shows that nonlinear optical pumping into the NV$^{0}$ quartet state ($^4\!A_2$) can explain NV center PL quenching. These measurements are the first quantitative experimental study of the $^4\!A_2$ state's role in dark state behavior, which is distinct from the non-fluorescent positive charge state (NV$^+$) studied by Fermi-level engineering\,\cite{Grotz2012, Pfender2017}. We demonstrate that this state allows imaging of local fields, establish an understanding of its photodynamics, and show that it allows optical switching\,\cite{Geiselmann2013}. Furthermore, we constrain the energy difference between $^4\!A_2$ and the NV$^0$ ground-state $^2\!E$, and the recombination energy threshold from $^4\!A_2$ to the NV$^-$ ground state $^3\!A_2$.  Understanding this behavior is critical for predicting the limits of sensors based on optical absorption\,\cite{Dumeige2013,Jensen2014,Chatzidrosos2017}, spin-optomechanical cavity\,\cite{Shandilya2021,Cady2019} and levitated nanoparticles\,\cite{Neukirch2013} systems, all of which use intense IR fields, as well as memories\,\cite{Dhomkar2016,Jayakumar2016} and other quantum devices\,\cite{Doherty2016} involving NV charge states\,\cite{Shields2015,Jaskula2019,Gulka2021,Zhang2021NatCom,Irber2021,Flagan2025}.

The diamond microdisk studied here is evanescently coupled to a fiber taper waveguide, and as shown in Fig.\,\ref{fig1}\,(a), positioned in a confocal microscope that excites NV centers in the device (NV concentration $\sim 10^{13}-10^{14}$\,cm$^{-3}$, Element Six optical grade material\,\cite{Acosta2009}) with a green laser. Room temperature photoluminescence is collected using both the microscope and the fiber taper, which select emission into free space and microdisk modes\,\cite{Masuda2024}, respectively. The fiber taper is also used to couple IR light (1500\,nm and 960\,nm wavelength bands) into microdisk modes. Intracavity IR field intensities $>$ 1\,W/$\upmu$m$^2$ for mW input power is possible thanks to the high-$Q$ and small mode volume of the cavity modes\,\cite{Coutts2026arXiv}. 

The photodynamics of NV centers excited by green and IR fields are illustrated in Fig.\,\ref{fig1}\,(b). The green laser causes NVs in NV$^0$ and NV$^-$ to cycle between their ground and excited states and emit PL. During this process, the green laser and IR fields modify the charge state by exciting electrons (holes) from the NV$^-$ (NV$^0$) excited (ground) state to diamond's conduction (valence) band. Central to our findings is that, at high IR power, two IR photons drive population from the $^3E$ state of NV$^-$ to the $^4\!A_2$ state of NV$^0$ in an ionization process nonlinearly dependent on IR field intensity.
This quenches PL of NVs positioned within the whispering gallery modes, and is vividly illustrated by comparing the change in PL intensity with and without the IR field as the microscope focus is rastered over the microdisk. The resulting images, shown in Fig.\,\ref{fig1}\,(c) for two different IR modes, clearly show that PL is suppressed near the microdisk's perimeter where the whispering gallery modes are localized.

\begin{figure}[!t] 
	\includegraphics[width=\linewidth]{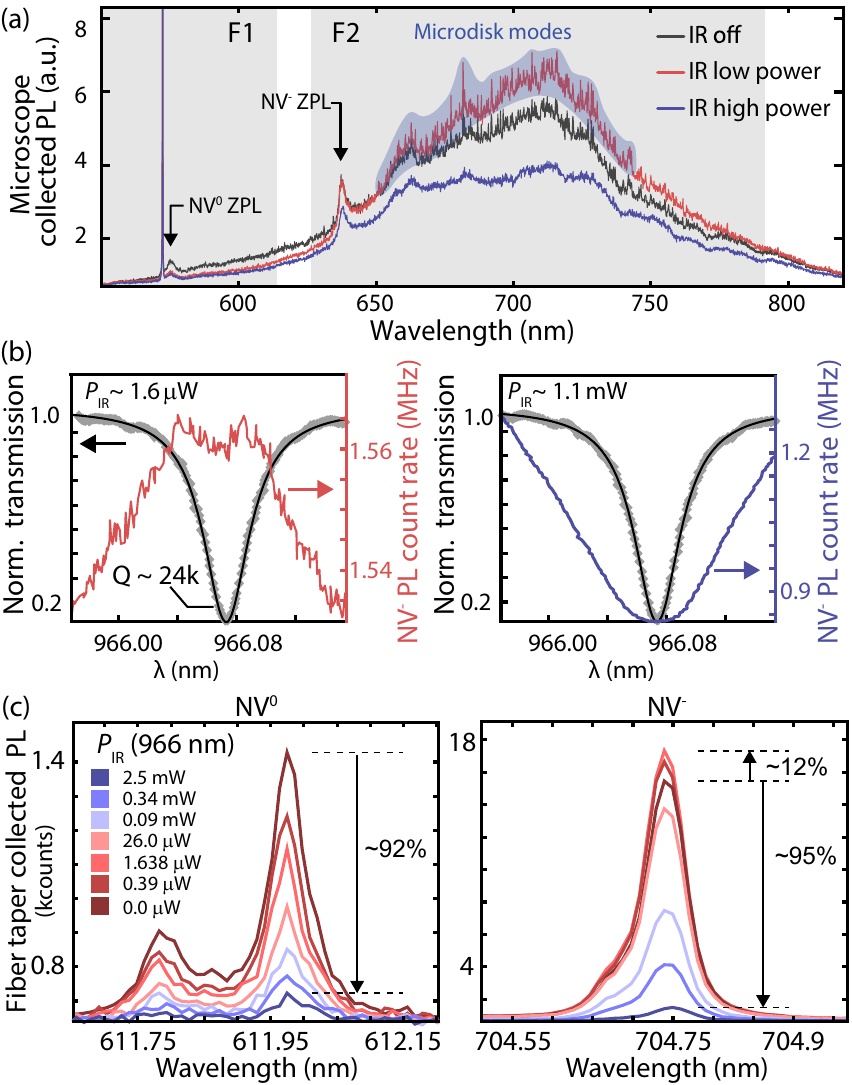}  
  \caption{
          \label{fig2}
     (a) Microscope measured PL spectra with the green laser focused on the microdisk edge for three IR (1524\,nm) field intensities input to the fiber taper: high (blue), low (red) and off (black). Gray shaded regions indicate the filter bands used to measure NV$^0$ and NV$^-$ emission strength. Sharp peaks in the blue shaded region are microdisk mode coupled PL. (b) Variation of NV$^-$ PL when the IR wavelength is tuned across a microdisk mode (in black) in the low (left, red) and high (right, blue) power regimes. (c) PL spectra obtained with fiber taper collection and $P_\text{G} = 4.6~\text{mW}$ for varying $P_\text{IR}$, showing emission into modes far within the NV$^0$ (left) and NV$^-$ (right) emission bands.  }
\end{figure}

The connection between quenching and change in NV charge state is revealed by changes to the PL spectrum. Figure\,\ref{fig2}\,(a) shows PL spectra collected by the microscope focused on the microdisk edge, for varying IR power, $P_\text{IR}$, input to the fiber taper and coupled to a 1524\,nm microdisk mode. Visible in each spectrum are the zero phonon lines of NV$^-$ (637\,nm) and NV$^0$ (575\,nm) and their phonon sidebands.  We observe that a weak IR field (red, $P_\text{IR}\sim1\,\upmu\text{W}$) enhances NV$^-$ PL (region F2: 627--793\,nm) and reduces NV$^0$ PL (region F1: 550--614\,nm), consistent with previous studies\,\cite{Ji2016}.
However, a stronger IR field (blue, $P_\text{IR}\sim1\,\text{mW}$) reduces PL from both NV$^0$ and NV$^-$, with NV$^-$ quenching being more pronounced. Similar behavior was observed for shorter wavelength IR light, as shown in Fig.\,\ref{fig2}\,(b), which plots NV$^-$ PL intensity as the IR wavelength is swept across a 966\,nm microdisk resonance. In these measurements, a bandpass filter spanning spectral region F2 followed by a single photon detector was used to integrate PL predominantly from NV$^-$. For $P_\text{IR} \sim 1\,\upmu\text{W}$, NV$^-$ PL increases with the IR laser on-resonance (Fig.\,\ref{fig2}\,(b), left). Conversely, for $P_\text{IR} \sim 1\,\text{mW}$, NV$^-$ PL decreases with the IR laser on-resonance (Fig.\,\ref{fig2}\,(b), right).  

Analyzing the IR field's influence on NV charge state is complicated by PL from  
the diamond substrate. This background can be eliminated using the fiber taper to selectively collect PL from NVs positioned within whispering gallery modes\,\cite{Masuda2024}. Figure\,\ref{fig2}\,(c) shows fiber taper collected PL spectra from modes within NV$^0$ (left) and NV$^-$ (right) emission bands. In both cases, PL is quenched by $> 90\,\%$  at the highest $P_\text{IR}$, significantly exceeding the $\sim 30\,\%$ quenching of the microscope spectra in Fig.\,\ref{fig2}\,(a). From this difference, the contribution from background NVs to microscope PL can be estimated and compensated for in subsequent measurements (see Supplemental Material, Sec. \Romannum{2}). Note the stronger quenching of cavity coupled PL is also evident from the suppression of sharp peaks in the microscope spectra in Fig.\,\ref{fig2}\,(a) originating from cavity mode PL\,\cite{Masuda2024}.

\begin{figure}[t!]
	\includegraphics[width=\linewidth]{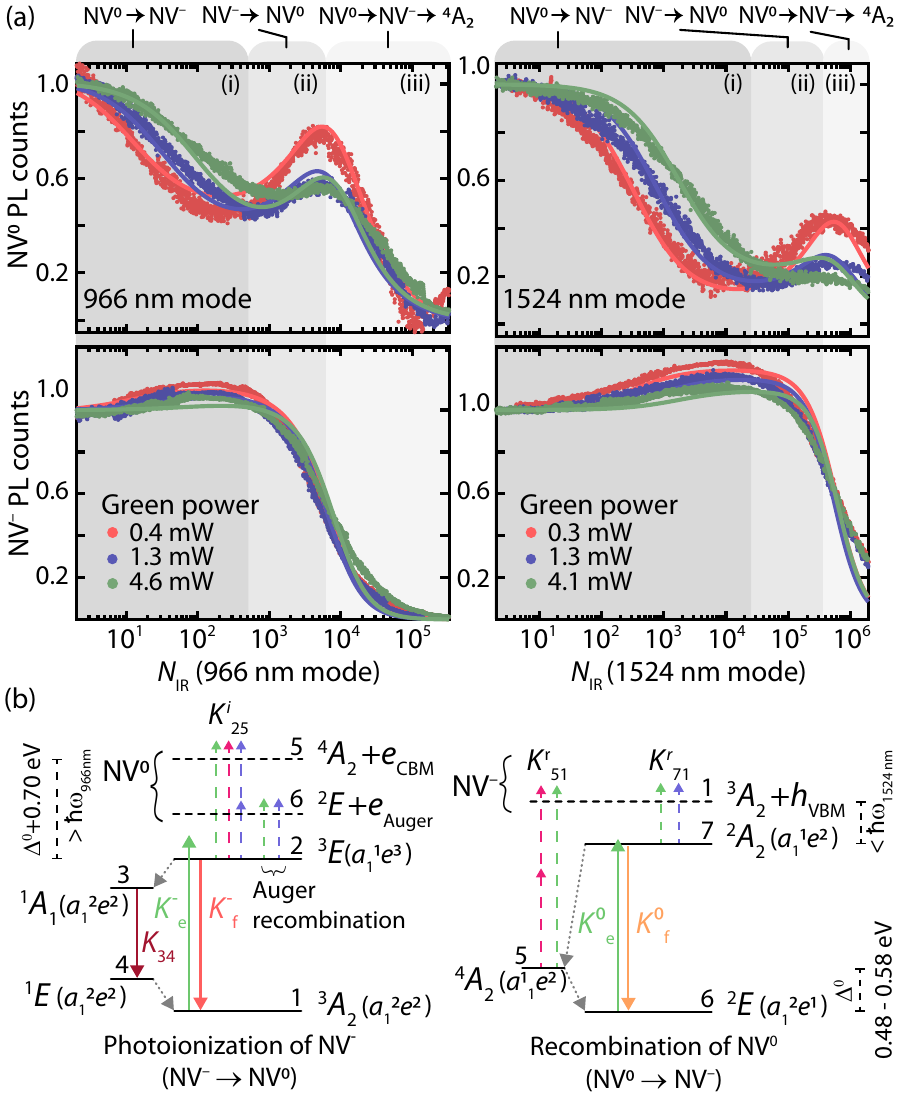}
		\caption{
		    \label{fig3}
              (a) Dependence of NV$^0$ and NV$^-$ PL on $N_\text{IR}$ of 966\,nm (left) and 1524\,nm  (right) microdisk modes for varying $P_\text{G}$. 
              Solid lines show fits derived from the model in (b), which shows the NV center states, transitions, and energy thresholds considered here. Straight dashed arrows indicate transitions between charge states. Blue (green) arrows indicate rates dependent on IR (green) field intensity. 
              Excitation rates $K^\text{i}_{25}$, $K^\text{i}_{26}$ and, $K^\text{r}_{71}$ depend on both IR and green field intensity. For 966\,nm, the rate $K^\text{r}_{51}$ depends on both IR and green field intensity, whereas for 1524\,nm, $K^\text{r}_{51}$ depends on green field intensity only.
              The pink arrows indicates transitions allowed for $966\,\text{nm}$ only. 
              The double blue arrow for $K^\text{i}_{25}$ indicates a two-photon process. Non-radiative transitions are indicated by dotted gray arrows.}
\end{figure}

A quantitative study of the influence of the IR field on NV charge state is shown in Fig.\,\ref{fig3}\,(a), which plots the dependence of NV$^0$ and NV$^-$ PL on intracavity IR photon number ($N_\text{IR}$).
This data was compiled by varying the IR field's power and detuning from resonance, allowing $N_\text{IR}$ to vary between $ \sim 1$ and $10^6$.
The predicted single photon field intensity is $\sim4\times10^{2}\,\text{W\,cm}^{-2}$ and $\sim1\times10^{2}\,\text{W\,cm}^{-2}$ for the $966\,\text{nm}$ and $1524\,\text{nm}$ modes, respectively (see Supplemental Material, Sec. \Romannum{3}). Emission from each charge state was measured by filtering the microscope PL in spectral regions F1 (NV$^0$) and F2 (NV$^-$), corrected for background PL, and normalized by the signal with no IR field. Figure\,\ref{fig3}\,(a) shows that NV$^0$ and NV$^-$ emission varies non-monotonically with $N_\text{IR}$ and respond qualitatively similar to IR fields at 966\,nm and 1524\,nm wavelengths, suggesting that a somewhat common mechanism underlies the charge state dynamics at both wavelengths. The observed behavior can be divided into three regions shown in Fig.\,\ref{fig3}\,(a). For low $N_\text{IR}$, NV$^0$ PL decreases and  NV$^-$ PL increases. For larger $N_\text{IR}$, this trend reverses. Finally, at high $N_\text{IR}$, PL from both charge states is quenched. Varying green laser power $P_\text{G} = $[0.4, 1.3, 4.6]\,mW, affects IR field driven NV ionization at low $N_\text{IR}$, but has no effect at high $N_\text{IR}$, as two IR photon pumping to $^4{A}_2$ dominates over all other dynamics.
Note that we did not observe any NV emission in absence of the green laser, as has been observed from multiphoton excitation of nanodiamonds\,\cite{Ji2018PRB}. 
 
To model the charge state dynamics, we follow Razinkovas et al.\,\cite{razinkovas2021photoionization} and consider the seven-level system in Fig.\,\ref{fig3}\,(b). Our model is limited to photoionization and recombination processes accessible with single green, single or two IR photon excitation (see Fig.\,S5 in the Supplemental Material, Sec. \Romannum{4}).
We disregard spin-dependent ionization, and assume that the system is initialized in the $m_\text{s}=0$ spin sub-level of the NV$^-$ orbital ground state by the green laser.
Our model permits one direct photoionization process with rate $K^i_{25}$, ionization via an Auger recombination process\,\cite{Gali2019,Wirtitsch2023} with rate $K^i_{26}$, and two recombination processes with rates $K^r_{51}$ and $K^r_{71}$. These rates can be decomposed into independent one and two-photon processes proportional to $N_\text{IR}$, and $P_\text{G}$ and $N_\text{IR}^2$, respectively (see Supplemental Material, Sec. \Romannum{5}):
\begin{align}
        K_{25}^i &= \bar{K}_{25,1-966\,\text{nm}}^{i}(N_\text{IR})+\bar{K}_\text{25,2-IR}^i (N_\text{IR}^2) +  \bar{K}_\text{25,1-G}^i (P_{\textrm{G}})\,
        \label{eq:K25},\\
        K_{26}^i&= \bar{K}_{26,1-\text{IR}}^i + \bar{K}_{26,1-\text{G}}^i\,
        \label{eq:K26},\\
        K_{51}^r &= \bar{K}_{51,2-966\,\text{nm}}^{r}(N_\text{IR}^2)+ \bar{K}_\text{51,1-G}^r (P_{\textrm{G}})\,
        \label{eq:K51},\\
        K_{71}^r &= \bar{K}_\text{71,1-IR}^r(N_\text{IR}) +  \bar{K}_\text{71,1-G}^r (P_{\textrm{G}}),
        \label{eq:K71}
\end{align}
where the indices label the levels in Fig.\,\ref{fig3}\,(b). $K^i_{25}$ is the ionization rate from NV$^-$ excited state $^3E$ to  NV$^0$ quartet state $^4\!A_2$, which decays non-radiatively to NV$^0$ ground-state $^2\!E$, and includes the absorption of one green photon ($K_\text{25,1-G}^i$), or one IR ($\bar{K}_\text{25,1-IR}^i$) or two IR photons ($\bar{K}_\text{25,2-IR}^i$).
However, as we elucidate below, $\bar{K}_\text{25,1-IR}^i$ is restricted to $966\,\text{nm}$ only.
$K^i_{26}$ denotes an indirect ionization process from $^3E$ to $^2\!E$ mediated by Auger recombination\,\cite{Gali2019,Wirtitsch2023}, as described in the Supplemental Material (see a5--$K^i_{26}$). From the $^3E$ state, Auger recombination can be induced by absorption of either a green photon or a single infrared photon at both the wavelengths considered here. 
For 966\,nm, the recombination rate $K^r_{51}$ from $^4\!A_2$ to $^3\!A_2$ depends on the absorption of one green or two IR photons. However, two-photon absorption is not energetically permitted for 1524\,nm. Therefore, under 1524\,nm illumination, this recombination process is restricted to the absorption of a green photon only.
$K^r_{71}$ is a recombination pathway from $^2\!A_2$ to $^3\!A_2$ and includes the absorption of an IR photon or a green photon.
We highlight that the green laser is necessary for excitation, photoionization, and recombination.

\begin{figure}[t!]
	\includegraphics[width=\linewidth]{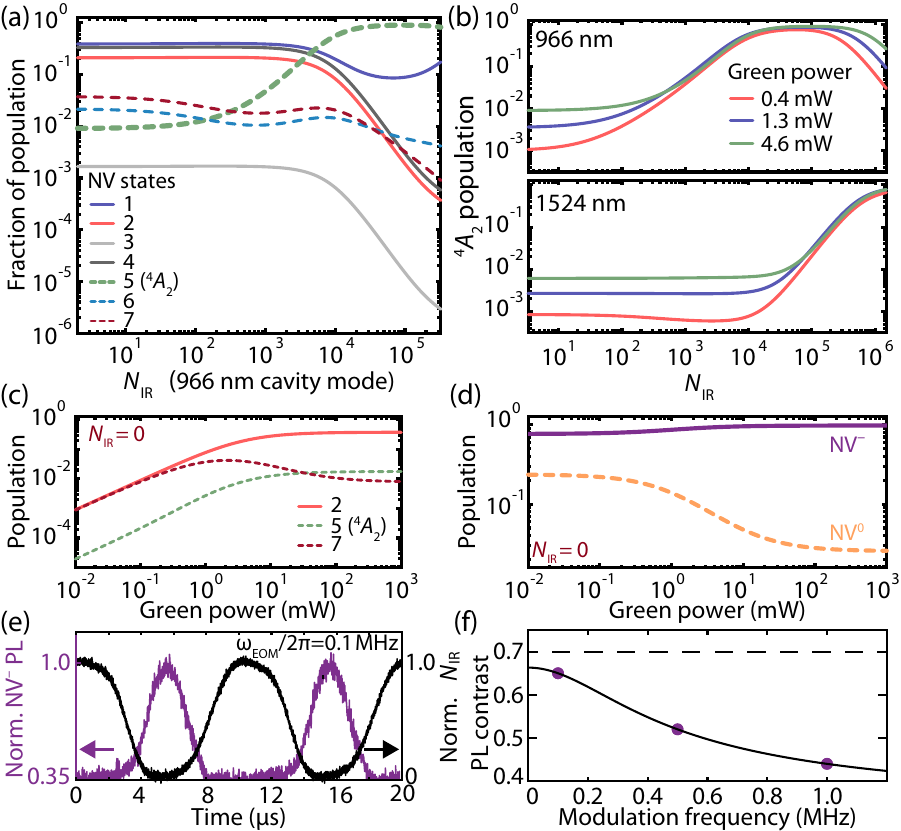}
		\caption{
		    \label{fig4}
	(a) Predicted dependence of NV state populations on $N_\text{IR}$ (966\,nm). Solid (dashed) lines represent NV$^-$ (NV$^0$) states.
        (b) Predicted change in population of the NV$^0$ quartet state ($^4$A$_2$) under 966\,nm (top) and 1524\,nm (bottom) excitation for various green laser powers. 
        Estimated (c) NV level population, and (d) relative population of NV$^-$ and NV$^0$, in absence of IR excitation. 
        (e) Time-resolved normalized NV$^-$ PL (purple line) when the 1524\,nm laser is modulated at $\omega_\text{EOM}/2\pi = 0.1\,\text{MHz}$. The black line shows the corresponding normalized $N_\text{IR}$. (f) Dependence of NV$^-$ PL contrast on modulation frequency. The dashed line shows the contrast expected from Fig.\,\ref{fig3}\,(a) as ${\omega_\text{EOM} \to 0}$ for 25\,dB extinction ratio.
     }
\end{figure}

Fitting this model to the PL data in Fig.\,\ref{fig3}\,(a) reproduces its dependence on IR and green field strength. The rates predicted by the fits for each component of $K_{25}^i$ and $K_{71}^r$ are summarized in Table\,\ref{tab:rates}. Examining the fits and the predicted steady-state populations, shown in Fig.\,\ref{fig4}\,(a), we see three regions of charge state dynamics. Region (i) is dominated by recombination of NV$^0$ to NV$^-$ via one-photon IR process ($K^r_{71}$, see Table\,\ref{tab:rates}), resulting in an increase of  NV$^-$ PL. In region (ii), two IR photon photoionization of NV$^-$ to NV$^0$ via  $^4\!A_2$ ($K^i_{25}$, see Table\,\ref{tab:rates}) becomes significant, resulting in an increase in NV$^0$ PL. Finally, in region (iii), the population of $^4\!A_2$ saturates, as population decay from $^4\!A_2$ to $^2\!E$ is limited by its constant non-radiative transition rate. This, combined with the continuous recombination of NV$^0$ to NV$^-$ via $K^r_{71}$, results in quenching of PL from both charge states.
Note that the predicted population of $^4\!A_2$, shown in Fig.\,\ref{fig4}\,(b), indicates that the 1524\,nm field in the experiment does not saturate the $^4\!A_2$ state at maximum $N_\text{IR}$.
The influence of the IR field is further illustrated by modeling the charge-state dynamics under green excitation only. Using the parameters extracted from the fits and $N_\text{IR} = 0$, the model predicts that the $^4\!A_2$ population increases monotonically and that the charge-state population remains largely unchanged with increasing green power, as shown in Figs.\,\ref{fig4}\,(c) and \ref{fig4}\,(d), respectively.

Based on our model we constrain the energies of some NV states (see Sec.\,\Romannum{4} in Supplemental Material for discussion).
First, the energy difference $\Delta^0  = [E(^4\!A_2)-E(^2\!E)]$ has been theoretically estimated as 0.48--0.68\,eV but has not been measured\,\cite{razinkovas2021photoionization,Ranjbar2011}.
Through analysis of the NV photoionization potential (IP), our experiments reduce this range. From the single-electron picture in Fig.\,S5 (Supplemental Material), we have that $\text{IP}(^3E\rightarrow{^4\!A_2})=\text{IP}(^3\!A_2 \rightarrow {^2\!E})-E^{\text{NV}^-}_\text{ZPL}+\Delta^0$, where IP$(^3\!A_2\rightarrow{^2\!E})=2.65$\,eV\,\cite{Aslam2013,Bourgeois2017} and $E^{\text{NV}^-}_\text{ZPL} = [E(^3E)-E(^3\!A_2)] = 1.946$\,eV\,\cite{Doherty2011}.
Our analysis shows that ionization $^3E \rightarrow {^4\!A_2}$ can be driven by a single 966\,nm photon ($\hbar\omega_{966\,\text{nm}} = 1.28\,\mathrm{eV}$), which is required to reproduce the low–IR-power data in Fig.\,\ref{fig3}\,(a). This constrains $\Delta^0 \leq 0.58\,\mathrm{eV}$, such that $E(^3E \rightarrow {^4\!A_2}) \leq 1.28\,\mathrm{eV}$ (see Sec.\,IV of the Supplemental Material).

Second, our model requires recombination from $^4\!A_2$ to $^3\!A_2$ to be allowed for a single green or two 966\,nm photons, i.e. $\text{R}(^4\!A_2\rightarrow{^3\!A_2})\leq\hbar\omega_{532\,\text{nm}}$, where R is the recombination threshold.
Consequentially, from $\text{R}(^4\!A_2\rightarrow{^3\!A_2}) \geq  \text{R}(^2\!A_2\rightarrow{^3\!A_2}) + E^{\text{NV}^0}_\text{ZPL} - \Delta^0$, we constrain $0.65\,\text{eV}<\text{R}(^2\!A_2\rightarrow{^3\!A_2})<0.75\,\text{eV}$.
Third, the recombination threshold from the NV$^0$ excited state $^2\!A_2$ is not well established, with Ji et al. suggesting ${\text{R}(^2\!A_2\rightarrow{^1\!E})<1.16\,\text{eV}}$\,\cite{Ji2016}. From  $\text{R}(^2\!A_2\rightarrow{^3\!A_2}) = \text{R}(^2\!A_2\rightarrow{^1\!E})- [E(^1\!E) - E(^3\!A_2)]$, we restricts $1.03\,\text{eV}<\text{R}(^2\!A_2\rightarrow{^1\!E})<1.13\,\text{eV}$ (see b1 and b2 in Sec.\,IV of the Supplemental Material). Here we have used $E(^1\!E)-E(^3\!A_2)$ = 0.38\,eV\,\cite{Bhandari2021}.
We refer to Fig.\,S5 in the Supplemental Material for a detailed description regarding the possible photoionization and recombination processes considered in this work. 

From the fits we can also estimate the $^4\!A_2$ state's effective decay rate, given by $K_{56} + K_\text{51}^r(P_\text{G})$. We find $K_{56}=165-562\,\text{kHz}$ for the intrinsic decay rate to the $^2\!E$ ground state and $K_\text{51,1-G}^r=438-1048\,\text{kHz/mW}$ for the recombination rate to the $^3\!A_2$ state of NV$^-$ (Fig.\,\ref{fig3}\,(b), and Table\,S3 of Supplemental Material, Sec.\,\Romannum{5}).
This estimate of $K_{56}$ corresponds to an intrinsic lifetime for $^4\!A_2$ of $\approx1.78-6.06\,\upmu\text{s}$, a value that to our knowledge has not previously been reported in the literature\,\cite{Felton2008, Baier2020, Thiering2024}.
Additional insight into $K_{56}$ can be obtained from the PL's response to a modulated IR field, which we implement using an electro-optic modulator with an extinction ratio of 25\,dB. Figure\,\ref{fig4}\,(e) shows time dependent NV$^-$ PL for $P_\text{G} = 4.1~\text{mW}$ when the 1524\,nm field is modulated at frequency $\omega_\text{EOM}/2\pi = 0.1$\,MHz, and Fig.\,\ref{fig4}\,(f) shows the effect of $\omega_\text{EOM}$ on PL contrast, defined by the maximum PL change induced by the IR field, normalized by the maximum PL with no modulation. For $\omega_\text{EOM}/2\pi = 0.1\,\text{MHz}$ the PL contrast is near what is expected from the DC measurements in Fig.\,\ref{fig3}(a) after accounting for the finite extinction of the modulator. For $\omega_\text{EOM}/2\pi = 0.5\,\text{MHz}$, the PL contrast decreases significantly.
This cutoff frequency is in good concordance with the effective decay rate of $2.1-4.2\,\text{MHz}$ predicted for $^4\!A_2$ from the fits for the value of $P_\text{G}$ used here.
Note that the modulator's 25\,dB extinction ratio correspond to a reduction in IR intensity of $N_\text{IR}\sim 10^6 \to 10^4$, where pumping from NV$^0$ to NV$^-$ will still be significant (Fig.\,\ref{fig3}\,(a)).
Further time-domain measurements and modeling are needed to better quantify the system's dynamics. In future, studying samples with tailored NV centers and nitrogen impurity density may reveal the influence of nitrogen impurities not explicitly considered in this work.

\renewcommand{\arraystretch}{1.2}
\begin{table}[t!]
\caption{Rates extracted from fitting experimental data with the model.}
\label{tab:rates}
\begin{tabularx}{.49\textwidth}{|c @{\extracolsep{\fill}} |c |c |c |}
\hline
Rate & 966\,nm & 1524\,nm & Unit\\
\hline
\hline
$K_\text{25, 1-G}^i$     & $35.11\pm14.30$                & $23.53\pm7.41$                 & kHz\,/\,mW \\ 
$K_\text{25, 1-IR}^i$    & $488.23\pm192.64$                & $-$                   & Hz\,/\,photon \\ 
$K_\text{25, 2-IR}^i$    & $0.20\pm0.08$   & $\left(3.47\pm1.07\right)\cdot10^{-5}$   & Hz\,/\,photon$^2$ \\ 
\hline
$K_\text{71, 1-G}^{r}$  & $19.62\pm8.86$                 & $8.19\pm2.78$                  & kHz\,/\,mW \\ 
$K_\text{71, 1-IR}^{r}$ & $1024.98\pm485.37$               & $27.94\pm9.79$                 & Hz\,/\,photon \\ 
\hline
\end{tabularx}
\end{table}

In summary, we have demonstrated modification to NV center PL by IR fields in a diamond cavity and shown that two-photon absorption can pump NVs into the dark $^4\!A_2$ state of NV$^0$ faster than the state's non-radiative decay, resulting in PL quenching of NV$^0$ and NV$^-$.   These results are crucial for understanding the $^4\!A_2$ state's impact on applications utilizing intense IR fields, such as nanoscopy\,\cite{Rittweger2009}, experiments with levitated diamonds\,\cite{Neukirch2013}, nonlinear optics\,\cite{Motojima2019,Flagan2025,Itoi2025arXiv}, and spin-optomechanics\,\cite{Shandilya2021}.
Of particular interest is quantum sensing schemes based on absorption of $1042\,\text{nm}$ photons by the NV$^-$ singlet state\,\cite{Chatzidrosos2017}. These schemes typically operate using large IR powers, where charge state switching can be expected to be significant, thus degrading the performance of these magnetometers. 
Furthermore, developing the ability to manipulate the $^4\!A_2$ state may enable applications based on its potentially long lifetime and spin lattice relaxation time\,\cite{Han2010, Felton2008}. Quenching based field imaging provides an alternative to electron beam\,\cite{Coenen2013}, nanoparticle\,\cite{Habteyes2014}, and near-field\,\cite{Rotenberg2014} techniques, and its nonlinear dependence on IR intensity may enable spatial resolution enhancement in post-processing. Our measurements also show that 1550\,nm light can implement STED super-resolution microscopy. Finally, because of the high penetration depth of IR light in biological tissues\,\cite{Helmchen2005}, IR quenching is promising for fluorescent contrast bioimaging\,\cite{Alkahtani2018} without background auto-fluorescence. 

\begin{acknowledgments}
We gratefully thank Viki Kumar Prasad and Matthias C. L{\"o}bl for fruitful discussions. We thank the anonymous Referee for their feedback regarding single photon ionization process. 
This work was supported by NSERC (Discovery Grant program), Alberta Innovates, and the Canadian Foundation for Innovation. SF acknowledges support from the Swiss National Science Foundation (Project No. P500PT\_206919).
\end{acknowledgments}


%

\end{document}


\title{\textit{Supplementary Material for}\\ Nonlinear optical charge state switching and pumping to a diamond NV center dark state}
\author{Prasoon K.\ Shandilya}
\thanks{These authors contributed equally.}
\author{Vinaya K.\ Kavatamane}
\thanks{These authors contributed equally.}
\author{Sigurd Fl{\aa}gan}
\thanks{These authors contributed equally.}
\author{David P.\ Lake}
\author{Denis Sukachev}
\author{Paul E.\ Barclay}

\email[Paul~E.\ Barclay:]{Corresponding author pbarclay@ucalgary.ca}
\affiliation{
Institute for Quantum Science and Technology, University of Calgary, Calgary, Alberta T2N 1N4, Canada}

\date{\today}
\maketitle
\section{Experimental Setup}
\begin{figure}[t!]
	\includegraphics[width=\linewidth]{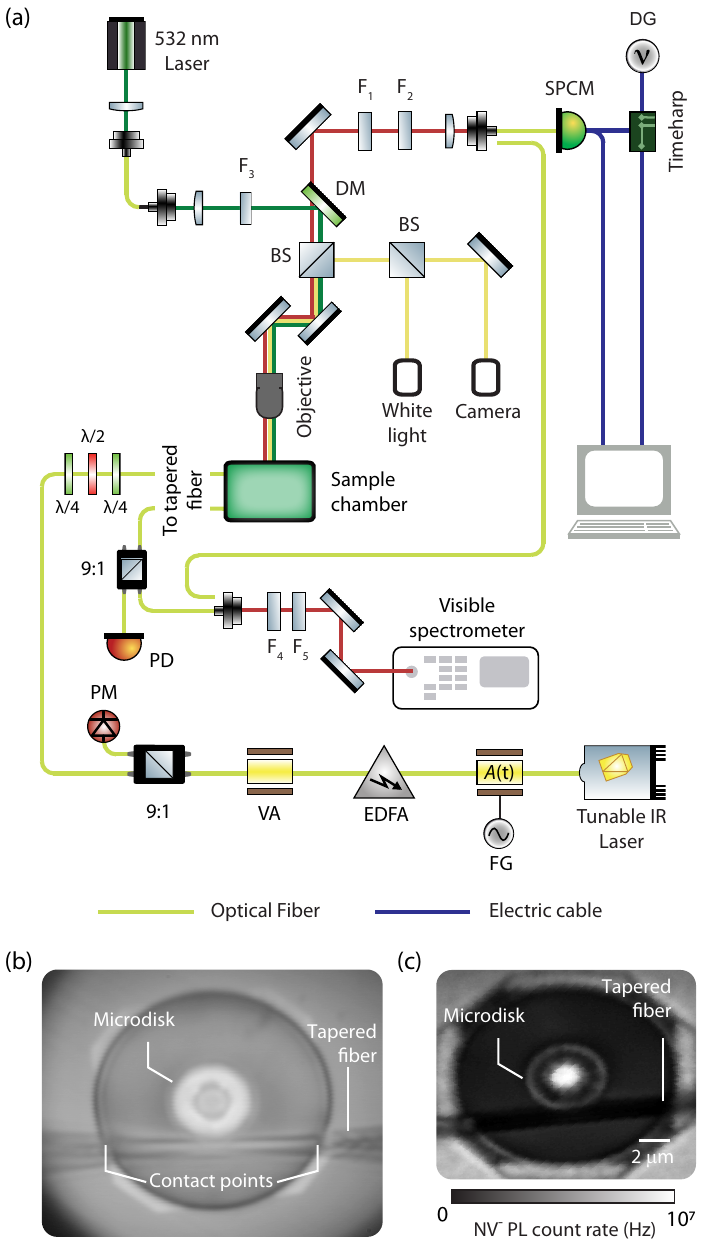}
		\caption{
		    \label{fig1}
		     (a) Schematic diagram of the experimental setup. Filter F3 is a 532\,nm notch filter, while filters F4 and F5 are shortpass filters used to block the IR lasers. See the main manuscript for the discussion of the other components.
      (b)  White light image of the diamond microdisk and the fiber-taper waveguide.
      (c) Confocal NV center PL raster scan of the system recorded under 532\,nm illumination.
       }
\end{figure}

\
In Fig.\,\ref{fig1}(a), we show a schematic diagram of the experimental setup used in this study. A diamond microdisk with diameter $d= 5.3\,\mu$m and thickness $h = 0.66\, \mu$m,  was patterned from a diamond chip (Element Six, optical grade) using quasi-isotropic plasma etching\,\cite{Mitchell2016,Mitchell2019}, and was previously used in recent spin-optomechanics experiments\,\cite{Shandilya2021}. To couple IR light into the diamond microdisk, we use tunable diode lasers (New Focus-6700 series) with 1520--1570\,nm and 940--985\,nm tuning ranges. To reach higher power levels, the 1520\,nm laser was amplified with an erbium-doped fiber amplifier (EDFA, Pritel LNHPFA-30). The input power was controlled using a variable attenuator (VA, Exfo FVA-3100) and monitored using the 10\,\% port of a 90:10 fiber beamsplitter connected to a power meter (PM, Newport Model 1936-R). For all of the measurements presented in the paper, a dimpled fiber taper used to couple light into and out of the microdisk was `parked’ on the shoulders of the etched circular trench surrounding the devices and in contact with the microdisk (see Fig.\,\ref{fig1}(b)). Touching the microdisk with the fiber taper degrades the quality factor of its optical resonances, however, it facilitates more efficient fiber taper collection of NV photoluminescence (PL) coupled into the microdisk modes\,\cite{Masuda2024}. Furthermore, contacting the microdisk also prevents the device from entering mechanical self-oscillations at high optical input power.  The fiber taper output was connected to a 90:10 fiber beamsplitter, with the 10\% port used to monitor the cavity transmission using a photodetector (Newfocus, 1623 IR Nanosecond photodetector) and the 90\% port spectrally filtered to remove the transmitted IR field before being sent to a spectrometer (Princeton Instruments Acton SP2750 with PIXIS 100B CCD detector) to measure the spectrum of light coupled into the fiber taper from the microdisk.

Confocal microscopy measurements of NV center PL were performed using a home-built setup. A 532\,nm laser (Crystal Laser CL532-025-SO) was used to excite the NV centers with a 100x objective lens (Nikon TU Plan ELWD, NA = 0.8) whose focal spot can be rastered across the sample using a 3D piezo nanopositioner (Piezosystem Jena Tritor 101). The PL collected by the objective was spectrally separated from the green and IR excitation using a dichroic mirror (DM, Chroma ZT532rdc-3) and appropriate bandpass filters, respectively, and was directed either to the spectrometer or to a single photon counting module (SPCM Excelitas, SPCM-AQ4C). NV$^0$ and NV$^-$ specific PL was separated using bandpass filters F1:\,550-614\,nm and F2:\,627-793\,nm, respectively.  During the measurements, the position of the confocal excitation spot on the microdisk was actively stabilized by periodically generating raster scanned PL images of the device  from the SPCM output as shown in   Fig.\,\ref{fig1}(c), and repositioning the objective as required to recenter the image.  

For the time-resolved measurement presented at the end of this Supplementary Information, the 1520\,nm laser was amplitude-modulated ($\mathcal{A}(t)$) with an electro-optic modulator (EOM, EOSpace AZ-0K5-10-PFA-SFA) driven by a function generator (FG, Agilent N5171B) while simultaneously recording a time-tagged PL histogram from the SPCM output. A delay generator (DG, Stanford Research Systems DG535) was used to sync the signal generator and a time-tagger (PicoQuant TimeHarp 260). To measure the time dependent power inside the cavity, the output of the fiber taper was sent to a high-bandwidth photoreceiver (Newport 1544B) and monitored using a digital oscilloscope (Tektronix DSA70804B, not shown in Fig.\,\ref{fig1}(a)).

\section{Fiber-coupled NV center PL}
\begin{figure}[t!]
	\includegraphics[width=\linewidth]{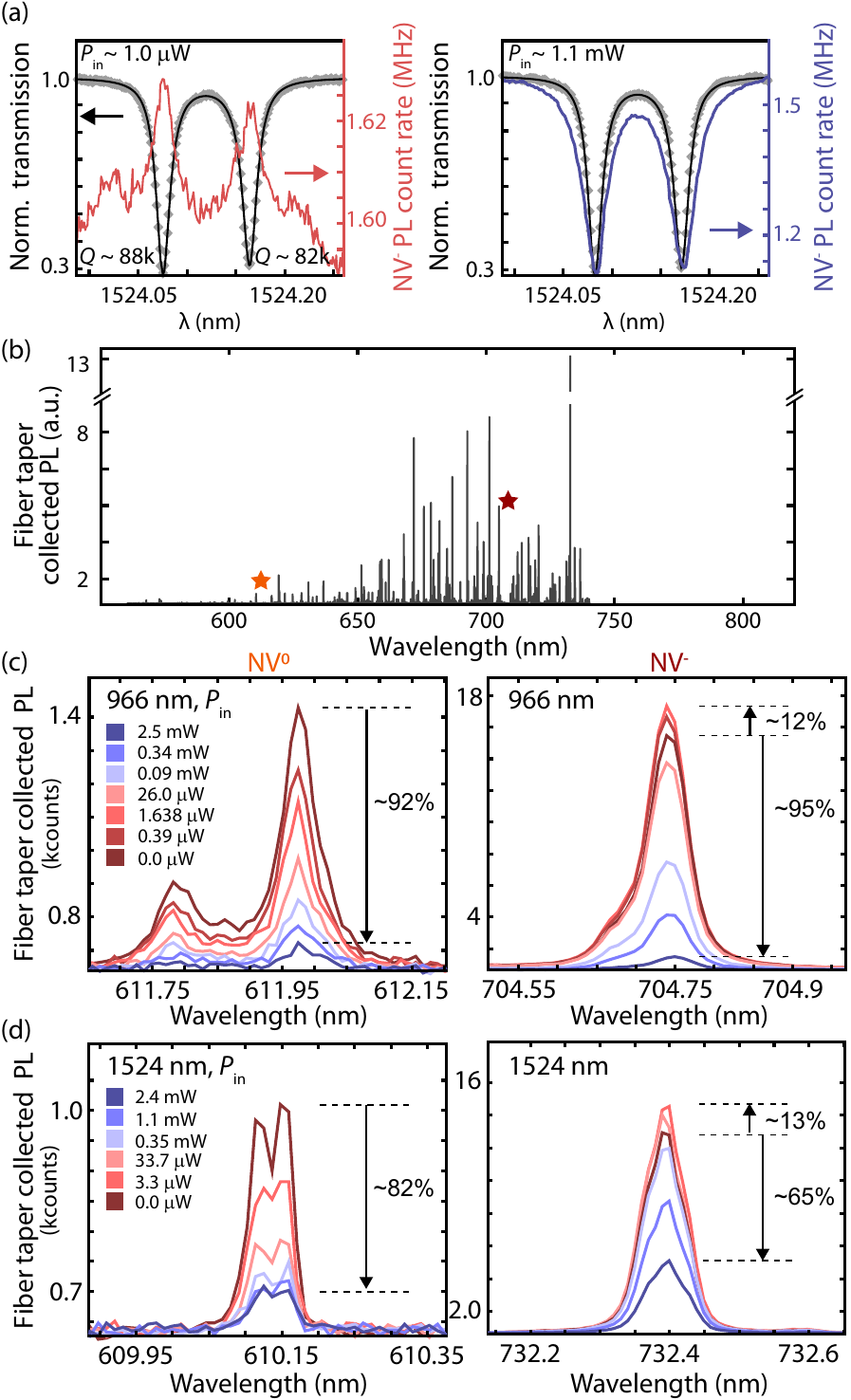}  
		\caption{
  \label{figPL}
    (a) Observed NV$^-$ PL with wavelength as the 1524\,nm laser is tuned across the cavity resonance (in black) in the low (left, red) and high (right, blue) power regime. Fitting the cavity transmission with a double Lorentzian yield loaded $Q\sim88\,000$ and $Q\sim82\,000$ for the doublet-mode.
    (b) Coupling between NV center PL and the whispering gallery modes detected via the fiber taper. The orange and red stars indicate the cavity modes used to explore PL emission from NV$^0$ and NV$^-$, respectively.
    (c) Whispering gallery mode coupled PL from NV$^0$ (left) and NV$^-$ (right) for varying 966\,nm laser power. For all measurements, the 966\,nm laser were kept on resonance with the cavity mode. Note that this panel is identical to as Fig.\,3(a) in the main manuscript. 
    (d) Same as panel (c) for the cavity mode at 1524\,nm mode.
     }
\end{figure}

In Fig.\,\ref{figPL}(a) we show the change in PL from NV$^-$ while scanning the 1520\,nm laser across the cavity resonance. The doublet nature observed for this cavity mode arises due to surface roughness induced back-scattering between degenerate clockwise and counter-clockwise propagating modes, resulting in non-degenerate standing wave modes\,\cite{Masuda2024, Borselli2004, Borselli2005}. Similar to Fig.\,2(b) of the main manuscript, we observe enhancement and quenching of the PL for $P_{\textrm{IR}}\sim\mu\text{W}$ and $P_{\textrm{IR}}\sim\text{mW}$, respectively. In these measurements, the contrast of the observed IR induced modification of the NV$^-$ PL is degraded by emission from NV centers positioned outside of the microdisk volume. For example, NV centers located below the microdisk but within the microscope objective's focal volume contribute to the spectra in Fig.\,2(a) in the main manuscript, but are unaffected by the IR field. This background PL can be accounted for by studying PL coupled into microdisk's whispering gallery modes since they are localized to a similar volume as the IR whispering gallery modes. To better isolate emission into these modes, we follow Ref.\,\cite{Masuda2024} and use the fiber taper to selectively collect their PL. The resulting spectrum is dominated by emission into cavity modes, as shown in Fig.\,\ref{figPL}(b). In Fig.\,\ref{figPL}(c), we show the influence of the 966\,nm cavity field on the NV PL emitted into whispering gallery modes at wavelengths corresponding to NV$^0$ ($< 612$\,nm) and NV$^-$ ($> 700$\,nm). We choose these wavelengths since NV$^-$ (NV$^0$) emission is negligible below 614\,nm (above 700\,nm)\,\cite{Rondin2010,Aslam2013}. We note that Fig.\,\ref{figPL}(c) is identical to Fig.\,2(c) of the main manuscript. With the 966\,nm laser on resonance  (see Fig.\,2(b) in the main manuscript), we observe a $\sim92\,\%$ reduction in NV$^0$ emission with increasing laser power. On the contrary, for NV$^-$ a $\sim12\,\%$ increase in emission is observed for low power, followed by a $\sim95\,\%$ quenching at high power. For completeness, in Fig.\,\ref{figPL}(d) we repeat the same measurement for the cavity mode at 1524\,nm. With the laser on resonance (Fig.\,\ref{figPL}(a)) we observe that emission from NV$^0$ is reduced by $\sim82\,\%$ with increasing IR power. Emission from NV$^-$ increases by $\sim13\,\%$ for low power, followed by a quenching of $\sim65\,\%$ at high power.

\section{NV-IR field intracavity interaction strength}

To analyze the interaction between NV centers and IR photons in the microdisk, we need to evaluate the effective intensity of the IR field inside the device. Here we present an analysis that takes into account the spatial field profiles of the microdisk modes at both IR and visible wavelengths, and their interaction with the volume of NVs excited by the green laser.

The power input to the fiber taper waveguide immediately before the coupling region with the microdisk is given by $P_\text{IR}$, and takes into account the fiber taper's non-ideal transmission efficiency, which is equal to $44\,\%$ and $52\,\%$ at the 1524\,nm and 966\,nm wavelengths used here, respectively. To generate plots showing charge state PL as a function of intracavity photon number,  $N_\text{IR}$, such as in Fig.\,3(a) of the main manuscript, data sets of PL vs.\ IR laser detuning from resonance, such as those in Fig.\,2(b) of the main manuscript and Fig.\,\ref{figPL}(a), were compiled for different $P_\text{IR}$. For each scan, the measured transmission spectrum was used to estimate $N_\text{IR}$ as a function of wavelength $\lambda = 2\pi c/\omega$ where $c$ is the speed of light and $\omega$ is the angular frequency of the input light. The mean intracavity photon number for a singlet (${N}^{s}_\text{IR}$) and doublet (${N}^{d}_\text{IR}$) mode of the microdisk cavity is given by\,\cite{Borselli2006}:
\begin{equation}
    \begin{split}
        {N}^{s}_\text{IR} &= \abs{\frac{\sqrt{\kappa_\text{ex}}}{\kappa/2- i\Delta}}^2 \frac{P_\text{IR}}{\hbar\omega}, \\
        {N}^{d}_\text{IR} &= \abs{\frac{\sqrt{\kappa_\text{ex}/2}}{\kappa/2-i(\Delta+\gamma_\beta/2)}+\frac{\sqrt{\kappa_\text{ex}/2}}{\kappa/2-i(\Delta-\gamma_\beta/2)}}^2 \frac{P_\text{IR}}{\hbar\omega}\,,
   \end{split}
\end{equation}
where $\Delta=\omega_\text{cav}-\omega$ is the cavity resonance--laser detuning, $\kappa$ is the total energy decay rate of the fiber coupled cavity mode, $\kappa_\text{ex}$ is the coupling rate from the cavity to the forward propagating fiber taper waveguide mode, and $\gamma_\beta$ is the internal back-scattering rate between clockwise and counter-clockwise propagating modes of the cavity. 

Estimating the intensity per photon in a microdisk or other integrated photonic cavity is often accomplished using an effective mode volume defined by the peak field strength. However, for the system studied here, we need to quantify the average field interacting with the distribution of NVs probed in the PL measurements. To account for this, we must consider how the optical process of interest depends on the IR field, and how efficiently the PL that is affected by this process is measured. 

The IR intracavity field drives an electronic transition with cross-section $\sigma$ at a rate whose dependence on field strength is governed by whether one or two photons are energetically required. Below we describe how to estimate coupling rates associated with single IR photons, before considering the more general case involving multiple IR photons.

\subsection{Single photon transitions}

For a one IR photon process, the transition rate for an NV positioned at coordinate $\mathbf{r}_i$ is,
\begin{align}
K_i = \frac{\sigma I_\text{IR}(\textbf{r}_i)}{\hbar\omega} = \frac{1}{2}\frac{\sigma}{\hbar\omega}\frac{c}{n_g}{\epsilon(\textbf{r}_i)\left|\mathbf{E}^\text{IR}_n(\mathbf{r}_i) \cdot \mathbf{u}_i\right|^2}, \label{eq:K1i}
\end{align}
where $I_\text{IR}(\textbf{r})$ is the spatially varying IR field intensity in the cavity\,\cite{Meirzada2018}. We relate this intensity to $\mathbf{E}^\text{IR}_n(\mathbf{r})$, the spatially varying electric field of the microdisk mode of interest, via the mode's group velocity, $c/n_g$, and the dielectric constant of the microdisk, $\epsilon$. Note that in optical cavities, the group index $n_g$ is related to the free spectral range of the cavity and will vary with mode order. In a microdisk, it is well approximated by the average refractive index sampled by the mode's energy density. Equation \eqref{eq:K1i} also accounts for misalignment between the NV dipole transition orientation, $\textbf{u}_i$, and the local field. 

In the PL based charge state measurements studied in the main manuscript, emission from many NVs is collected. Assuming that PL is monitored through fiber taper collection of emission into a single cavity mode, the average $K$ of NVs participating in these measurements is given by:
\begin{equation}
\left\langle K \right\rangle = \frac{1}{N_\text{col}}\sum_i \alpha_{e,i}^\text{532\,nm}\alpha_{m,i}^\text{NV} \eta_m K_i 
\end{equation}
where $\alpha_{e,i}^\text{532\,nm}$ is the probability of the NV at position $\mathbf{r}_i$ being excited by the 532\,nm laser, $\alpha_{m,i}^\text{NV}$ is the probability of the resulting NV emission into its phonon sideband coupling to mode $m$ of the microdisk that is being monitored in a given measurement, and $\eta_m$ is the probability of photons in this mode being collected by the fiber taper waveguide.  This expression is normalized by the total number of photons collected by the measurement, 
\begin{equation}
    N_\text{col} = \sum_i \alpha_{e,i}^\text{532\,nm}\alpha_{m,i}^\text{NV} \eta_m.
\end{equation}
The excitation probability is proportional to the green laser intensity profile created by the microscope, $\alpha_{e,i}^\text{532\,nm} = \alpha_{e,0}^\text{532\,nm}|\mathbf{E}^\text{532\,nm}(\mathbf{r}_i)|^2$, which in general will depend on the green laser power, the microscope spot size, and optical scattering of the microscope spot by the microdisk structure. Owing to the Purcell effect, the probability of the resulting NV phonon sideband emission coupling into mode $m$ of the microdisk is proportional to the mode's field profile, $\alpha_{m,i}^\text{NV} = \alpha_{m,0}^\text{NV}|\mathbf{E}^\text{NV}_m(\mathbf{r}_i)|^2$. In each of these expressions we have assumed that the likelihood of the relevant dipole matrix element being aligned with the electric field has been captured by the constant pre-factor. Combining these expressions, we can express the average excitation rate as:
\begin{equation}
    \left\langle K \right\rangle = \frac{\sigma c}{2 \hbar\omega n_g}\frac{\sum_i|\mathbf{E}^\text{532\,nm}(\mathbf{r}_i)|^2 |\mathbf{E}^\text{NV}_m(\mathbf{r}_i)|^2 \epsilon(\mathbf{r}_i)\left|\mathbf{E}^\text{IR}_n(\mathbf{r_i}) \cdot \mathbf{u}_i\right|^2}{\sum_i|\mathbf{E}^\text{532\,nm}(\mathbf{r}_i)|^2 |\mathbf{E}^\text{NV}_m(\mathbf{r}_i)|^2}.
\end{equation}
Rewriting this as an integral over the microdisk volume
\begin{equation}
    \left\langle K \right\rangle = \frac{\overline{\sigma} c \epsilon_\text{dia}}{2 \hbar\omega n_g}\frac{\int_{\text{md}} |\mathbf{E}^\text{532\,nm}(\mathbf{r})|^2 |\mathbf{E}^\text{NV}_m(\mathbf{r})|^2 |\mathbf{E}^\text{IR}_n(\mathbf{r})|^2 d^3\mathbf{r}}{\int_{\text{md}}|\mathbf{E}^\text{532\,nm}(\mathbf{r})|^2 |\mathbf{E}^\text{NV}_m(\mathbf{r})|^2 d^3\mathbf{r}},
\end{equation}
where we have accounted for the average misalignment between the field in IR mode $n$ and the NV dipole transition of interest with $\overline{\sigma} \le \sigma$, and $\epsilon_\text{dia}$ is the dielectric constant of diamond. Noting that the intracavity energy stored in the IR field is $U = \int d\text{r}^3 \epsilon |\mathbf{E}^\text{IR}_n|^2$/2, this expression can be rewritten as follows
\begin{equation}
    \left\langle K \right\rangle = \overline{\sigma} \frac{ c}{ \hbar\omega n_g} \frac{U}{V_o^\text{IR}}\Gamma = \overline{\sigma} \frac{ c}{ n_g} \frac{N_\text{IR}}{V_o^\text{IR}}\Gamma \label{eq:K_single}
\end{equation}
where $V_o^\text{IR} = \int\epsilon(\mathbf{r})|\mathbf{E}^\text{IR}_n(\mathbf{r})|^2 d^3\mathbf{r}/|\epsilon\mathbf{E}_n^\text{IR}|^2_\text{max}$ is IR field's mode volume defined by its peak energy density, and
\begin{align}
    \Gamma = \frac{\int_{\text{md}}{|\mathbf{E}^\text{IR}_n(\mathbf{r})|}^2|\mathbf{E}^\text{532\,nm}(\mathbf{r})|^2 |\mathbf{E}^\text{NV}_m(\mathbf{r})|^2  d^3\mathbf{r}}{|\mathbf{E}_n^\text{IR}|^2_\text{max}\int_{\text{md}}|\mathbf{E}^\text{532\,nm}(\mathbf{r})|^2 |\mathbf{E}^\text{NV}_m(\mathbf{r})|^2 d^3\mathbf{r}} \label{eq:Gamma}
\end{align}
is a dimensionless confinement factor that accounts for the overlap between the intensity profiles of the IR mode, the green excitation spot, and the mode into which the NV emission is being collected. Precisely evaluating $\Gamma$ for a given measurement is complicated by uncertainty in the profile of the green laser excitation spot, as well as identifying which modes $m$ and $n$ the IR input field and NV emission are coupled to, respectively. 

To estimate $\Gamma$, we can simplify Eq.\ \eqref{eq:Gamma} by assuming that the green laser field is constant within the excitation spot volume, and zero outside of this volume, giving:
\begin{align}
    \Gamma = \frac{\int_{\text{ex}}{|\mathbf{E}^\text{IR}_n(\mathbf{r})|}^2 |\mathbf{E}^\text{NV}_m(\mathbf{r})|^2  d^3\mathbf{r}}{|\mathbf{E}_n^\text{IR}|^2_\text{max}\int_{\text{ex}} |\mathbf{E}^\text{NV}_m(\mathbf{r})|^2 d^3\mathbf{r}}.
\end{align}
This shows that $0 \le \Gamma \le 1$ is given by the mean normalized IR field intensity within the green laser spot, weighted by its overlap with the microdisk mode into which the NV phonon sideband emission is being monitored. This expression reaches a maximum values $\Gamma \to 1$ in the limit that the excitation spot is tightly focused where the IR mode intensity is maximum. Conversely, $\Gamma \to 0$ if the excitation spot or the NV mode do not overlap with the IR mode.

\subsection{Multi-photon transitions}
Transition rates involving multiple IR photons can be written as,
\begin{align}
    K^{(p)} &= \frac{\sigma^{(p)} I_\text{IR}^p(\textbf{r}_i)}{(\hbar\omega)^p} \\ 
    &= \sigma^{(p)}\frac{ c^p }{(2\hbar\omega n_g)^p}{\epsilon(\textbf{r}_i)^p\left|\mathbf{E}^\text{IR}_n(\mathbf{r}_i) \cdot \mathbf{u}_i\right|^{2p}},
\end{align}
where $\sigma^{(p)}$ is the $p$-photon cross-section of the transition of interest\,\cite{Higbie2017}. Following a similar procedure as for the single photon case, we find that 
\begin{equation}
    \left\langle K^{(p)} \right\rangle = 
     \overline{\sigma}^{(p)}\left(\frac{c}{ n_g} \frac{N_\text{IR}}{V_o^\text{IR}}\right)^p\Gamma^{(p)} \label{eq:K_multi}
\end{equation}
with 
\begin{equation}
        \Gamma^{(p)} = \frac{\int_{\text{md}}{|\mathbf{E}^\text{IR}_n(\mathbf{r})|}^{2p}|\mathbf{E}^\text{532\,nm}(\mathbf{r})|^2 |\mathbf{E}^\text{NV}_m(\mathbf{r})|^2  d^3\mathbf{r}}{(|\mathbf{E}_n^\text{IR}|_\text{max})^{2p}\int_{\text{md}}|\mathbf{E}^\text{532\,nm}(\mathbf{r})|^2 |\mathbf{E}^\text{NV}_m(\mathbf{r})|^2 d^3\mathbf{r}} \label{eq:Gamma_multi}.
\end{equation}
These equations show that the multi-photon transition rate depends upon the overlap of the $p$'th power of the electric field energy density with both the green laser excitation spot and the mode used to collect the NV emission.

\subsection{Numerical estimates of modal coupling parameters}

As discussed above, precisely predicting the cavity related parameters in Eqs.\ \eqref{eq:K_single} and \eqref{eq:K_multi} is complicated by several factors. The microdisk supports a complex spectrum of modes in both the IR and NV wavelength ranges, and identifying them with certainty is a challenge. Both the mode volume, $V_o$, and group index, $n_g$, vary between whispering gallery mode families. Modes that are fundamental in the radial and vertical directions typically have the smallest $V_o$ and largest $n_g$. Uncertainty in mode identification also complicates calculation of overlap factors $\Gamma^{(p)}$, which depend on the overlap of the IR and NV modes. Further uncertainty arises from the intensity distribution of the green laser field, as the nominally Gaussian green laser spot created by the microscope is scattered by the microdisk surfaces, resulting in a potentially complex intensity distribution.

\begin{figure}[t!]
	\includegraphics[width=\linewidth]{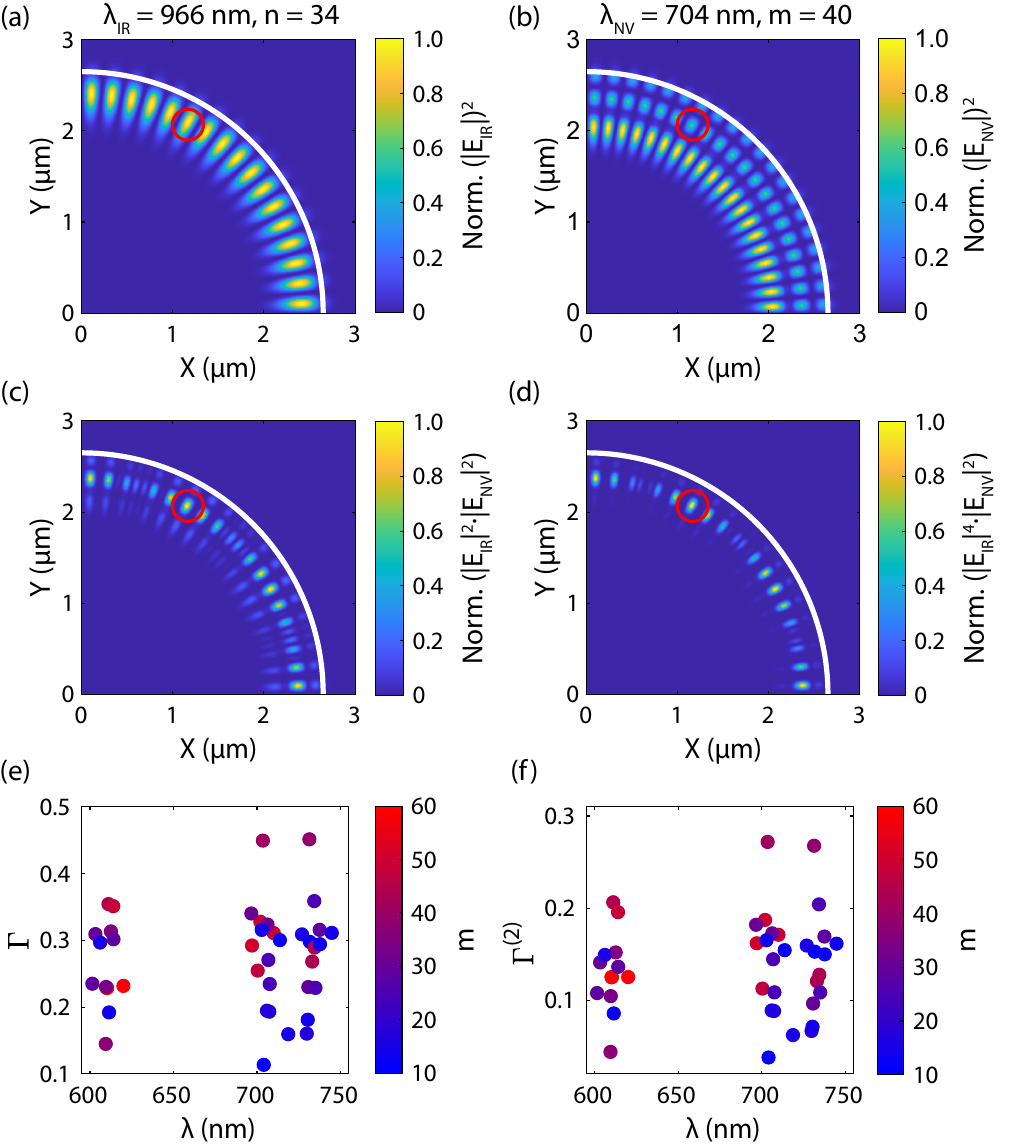}
		\caption{FEM simulations of the diamond microdisk. The white arch indicates the physical extent of the microdisk, whereas the red circle indicates the spatial extent of the spot-size of the green excitation laser. The excitation spot is placed on a maximum of the overlap between the IR and NV fields.
        (a) Normalized electric field profile, $|\mathbf{E}_n^\text{IR}|^2$, for the $966\,\text{nm}$ fundamental cavity mode with azimuthal mode number $n=34$.
        (b) Example of a the normalized electric field profile, $|\mathbf{E}_m^\text{NV}|^2$, for a higher-order mode at $704\,\text{nm}$, with azimuthal mode number $m=40$.
        (c) Normalized product of the fields $|\mathbf{E}_n^\text{IR}|^2$ and $|\mathbf{E}_m^\text{NV}|^2$ used to calculate $\Gamma=0.45$. 
        (d) Normalized product of $|\mathbf{E}_n^\text{IR}|^{2p}$ and $|\mathbf{E}_m^\text{NV}|^2$ for $p=2$, used in the calculation of $\Gamma^{(2)}=0.27$.
        (e) Dependence of $\Gamma$ with wavelength and $m$ for the $966\,\text{nm}$ fundamental cavity mode.
        (f) Dependence of $\Gamma^{(2)}$ with wavelength and $m$ for the $966\,\text{nm}$ fundamental cavity mode.
      }   \label{fig:mode_overlap}
\end{figure}

With these limitations in mind, we have performed calculations, using COMSOL finite element method (FEM) simulations, aiming to estimate these parameters (Fig.\,\ref{fig:mode_overlap}). As summarized in Table \ref{table_simulation}, we have calculated $V_o$ and $n_g$ for the fundamental TE modes near the IR wavelengths used in the measurements (Fig.\,\ref{fig:mode_overlap}\,(a)). For each of these modes, we have then calculated $\Gamma$ and $\Gamma^{(2)}$ for several candidate NV wavelength modes (Fig.\,\ref{fig:mode_overlap}). These calculations are performed assuming that the green laser spot can be approximated as constant within a cylinder whose diameter is equal to the full width at half max of the microscope field intensity and is aligned with the maximum of the IR field.

We show the results of numerical calculations of $\Gamma$ and $\Gamma^{(2)}$ for the $966\,\text{nm}$ cavity mode in Fig.\,\ref{fig:mode_overlap}\,(e) and (f), respectively. We performed the calculations for a variety of NV modes whose wavelengths are close to the modes investigated in Fig.\,\ref{figPL}\,(c)-(d). We observe that both $\Gamma$ and $\Gamma^{(2)}$ increases with increasing $m$ of the NV mode. {For an approximately fixed wavelength, increasing $m$ corresponds to reducing a mode's radial and vertical order, with lower order modes typically having stronger confinement.} For the range of wavelengths and values of $m$ investigated here, we find $0.11<\Gamma<0.45$ and $0.03<\Gamma^{(2)}<0.27$ for the $966\,\text{nm}$ IR mode. Similarly, for the $1524\,\text{nm}$ IR mode, we find $0.17<\Gamma<0.52$ and $0.06<\Gamma^{(2)}<0.33$ (not shown). 

\subsection{IR field intensity}

The intensity per photon can be calculated from Eq.\,\ref{eq:K1i}:
\begin{equation}
    I_{\text{IR}}=\frac{1}{2}\frac{c}{n_g}\epsilon\abs{\mathbf{E}^\text{IR}_n}^2\,,
\end{equation}
from which the maximum electric field per photon is given by\,\cite{Khitrova2006,Flagan2022} 
\begin{equation}
\abs{\mathbf{E}^\text{IR}_n}_\text{max}=\sqrt{\frac{\hbar\omega}{2\epsilon V_o^\text{IR}}}\,,
\end{equation}
when it is located within the diamond microdisk and 
\begin{equation}
    n_g = c\frac{\partial k_m}{\partial \omega_m}\,.
    \label{eq:n_g}
\end{equation}
Here, $k_m = m/R_{\text{eff}}$ is the wavenumber for the cavity mode with azimuthal mode number $m$ and $R_{\text{eff}}$ is an effective radius given by\,\cite{Coutts2026arXiv}
\begin{equation}
       R_{\text{eff}} = \frac{\int r\epsilon(\mathbf{r})|\mathbf{E}(\mathbf{r})|^2d^3\mathbf{r}}{\int\epsilon(\mathbf{r})|\mathbf{E}(\mathbf{r})|^2d^3\mathbf{r}}\,.
   \label{eq:Reff}
\end{equation}
By substituting $k_m$ and Eq.\,\ref{eq:Reff} into Eq.\,\ref{eq:n_g}, we obtain\,\cite{Coutts2026arXiv}
\begin{equation}
    n_{\text{g}} = \frac{c}{R_{\text{eff}}\frac{\partial\omega}{\partial m}} - \frac{cm}{R_{\text{eff}}^2}\frac{\partial R_{\text{eff}}}{\partial \omega}\,.
    \label{eq:groupIndex2}
\end{equation}

Using the values listed in Table\,\ref{table_simulation}, we find $I_{966\,\text{nm}}=3.57\times10^{6}\,\text{W\,m}^{-2}$ and $I_{1524\,\text{nm}}=1.30\times10^{6}\,\text{W\,m}^{-2}$. In Fig.\,3 of the main manuscript, we load the cavity with ${\sim\!10^5}$ and ${\sim\!10^6}$ photons for the $966\,\text{nm}$ and $1524\,\text{nm}$ cavity mode, respectively, which translates to a maximum field intensity of 
 $I_{966\,\text{nm}}=3.57\times10^{11}\,\text{W\,m}^{-2}$ and $I_{1524\,\text{nm}}=1.30\times10^{12}\,\text{W\,m}^{-2}$ for the respective cavity modes.


\renewcommand{\arraystretch}{1.5}
\begin{table}[!tb]
\centering
\caption{Key figures of merit obtained from the FEM simulations. 
\label{table_simulation}}
\begin{tabularx}{.49\textwidth}{|c @{\extracolsep{\fill}} cccccc|}
\hline
\makecell{Cavity\\ mode\\ (nm)} & $m$ & \makecell{$\frac{V_o^\text{IR}}{(\lambda/n_\text{dia})^3}$} & $n_g$ & $\Gamma$ & \makecell{$\Gamma^{(2)}$} & \makecell{$I_{\text{IR}}$ \\ ($\text{Wm}^{-2}$)} \\
\hline
\hline
966 & 34 & 29 & 2.42 & 0.11--0.45 & 0.03--0.27 & $3.57\times10^{6}$ \\ 
\hline
\hline
1524 & 19 & 12 & 2.59 & 0.17--0.52 & 0.06--0.33 & $1.30\times10^{6}$ \\
\hline
\end{tabularx}
\end{table}

\section{NV center photoionization and recombination processes}
\begin{figure}[t!]
	\includegraphics[width=\linewidth]{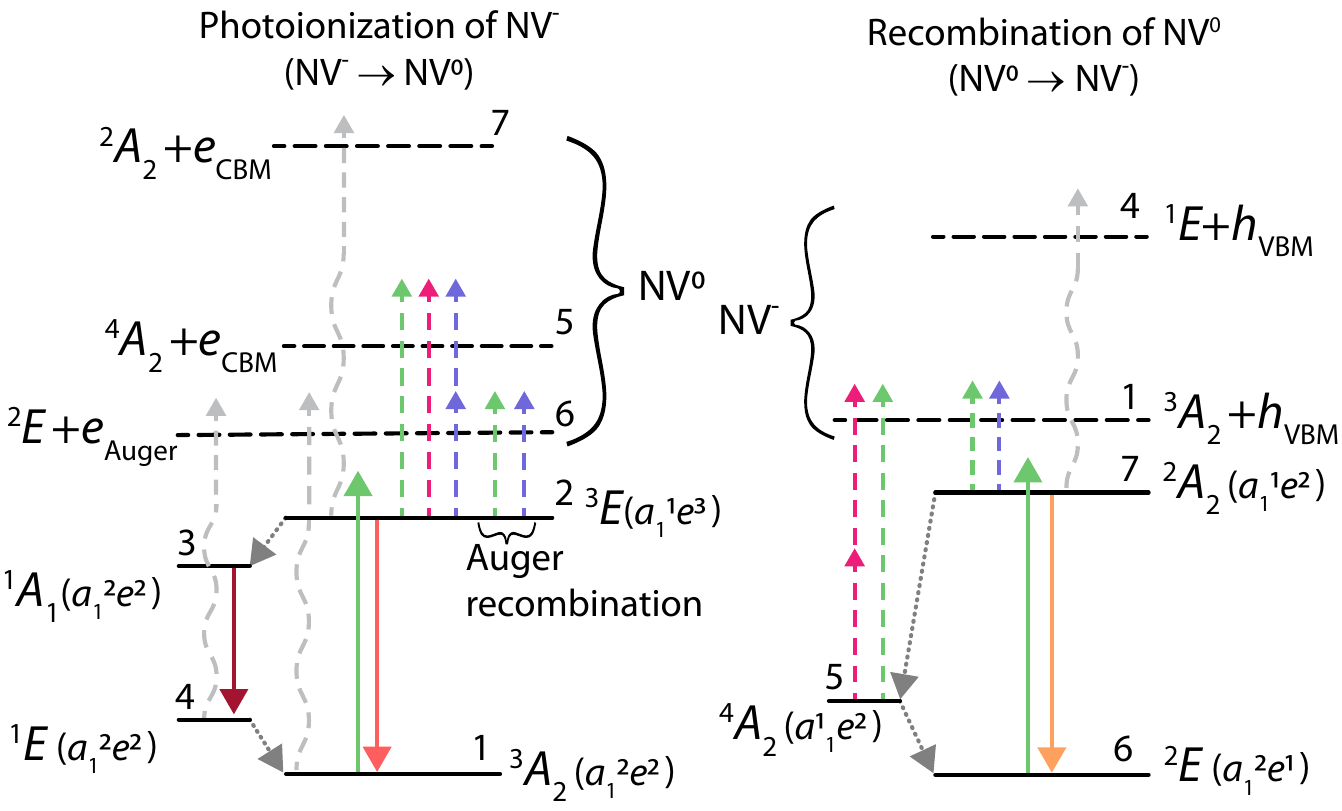}
		\caption{
		    \label{fig:model}
            Schematic diagram of the NV center energy levels with theoretically allowed photoionization (left) and recombination (right) processes. 
            The purple arrows indicate transitions allowed for both $966\,\text{nm}$ and $1524\,\text{nm}$, while the pink arrows indicates transitions that are allowed for $966\,\text{nm}$ only. Green arrows represents green photons.
            The straight dashed arrows indicate transitions involving a change in charge state, straight solid arrows represent electronic transitions within the same charge state, straight dotted gray arrows indicate non-radiative transitions, while the wavy dashed gray arrows indicate processes that are not supported in our system.
            $e_{\text{CBM}}$ and $h_{\text{VBM}}$ are energies required to excite an electron to the conduction-band minimum and a hole to the valence-band minimum, respectively.}
\end{figure}

As discussed in the main manuscript, we analyze the energy thresholds for ionization and recombination processes by following the description of the NV center's electronic states presented by Razinkovas et al. in Ref.\,\cite{razinkovas2021photoionization}. A threshold for photoionization corresponds to the energy required to excite an electron to the conduction-band minimum (CBM) of the diamond crystal lattice. Similarly, for recombination, the threshold corresponds to the energy required to excite a hole to the valence-band minimum (VBM) of the diamond crystal lattice\,\cite{razinkovas2021photoionization}. Photoionization and recombination processes can be depicted using the Jablonski diagram of the entire system shown in Fig.\,\ref{fig:model}. Here, photoionization (left) is represented by the NV$^-$ and NV$^0$ energy levels with an electron at the CBM ($e_{\text{CBM}}$), while recombination (right) is represented by the energy levels for NV$^-$ and NV$^0$ with a hole at the VBM ($h_{\text{VBM}}$).

\begin{figure*}[t!]
\includegraphics[width=\linewidth]{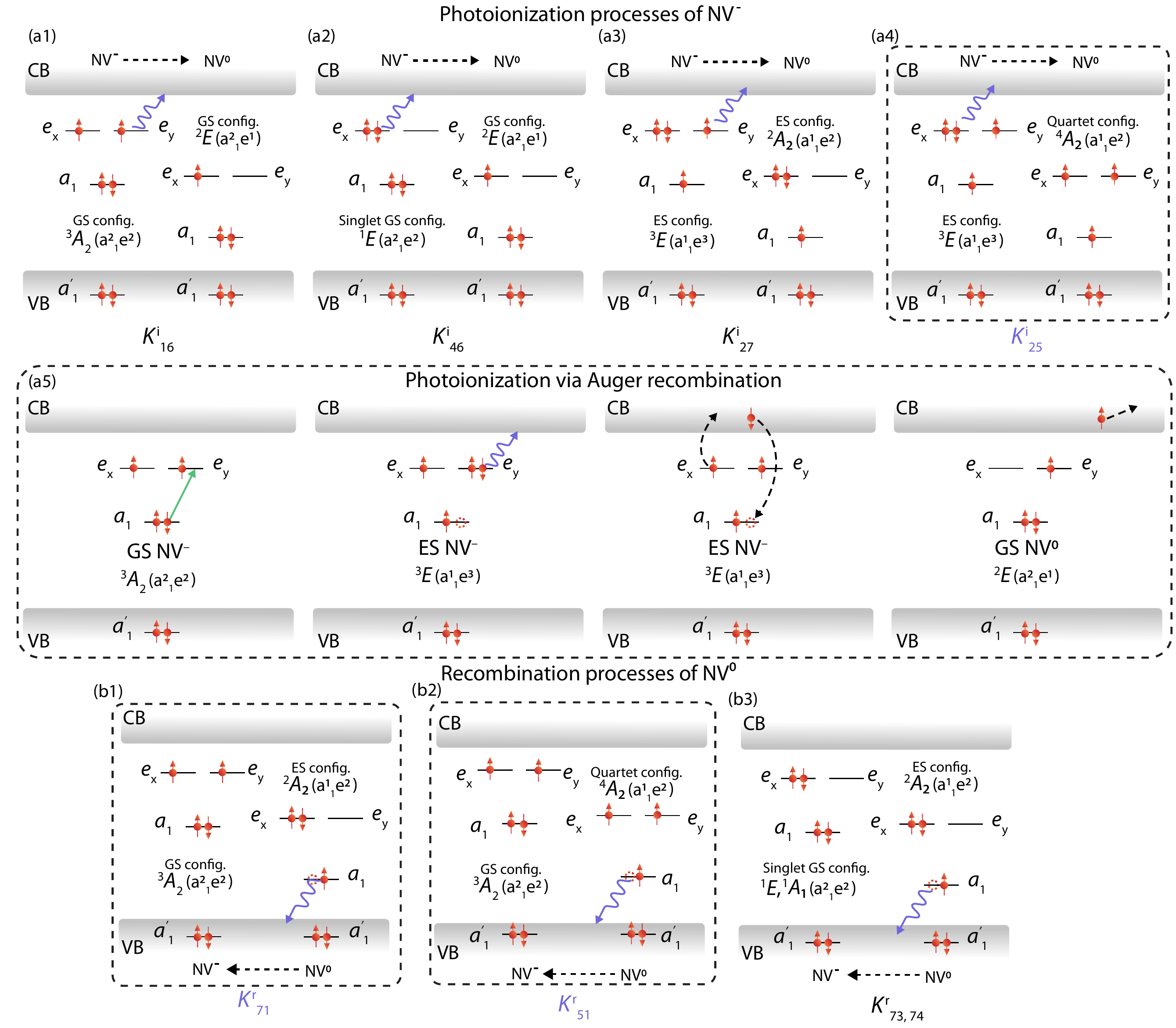}
		\caption{
		    \label{fig:electronModel}
             Single-electron picture of the NV center. Wavy arrows indicate possible photoionization and recombination processes. The processes highlighted by the dashed boxes form the basis of our study. (a) Available photoionization processes, NV$^-\rightarrow$ NV$^0$.
            (a1) $K^i_{16}$: Electronic configuration of the $m_s$ = 0 spin sublevel of the $^3\!A_2$ NV$^-$ ground state ionizing to the $m_s$ = 1/2 spin sublevel of $^2\!E$ ground state of NV$^0$.
            (a2)  $K^i_{46}$: Electronic configuration of the ground state singlet spin level $^1\!E$ of the NV$^-$ state ionizing to $m_s$ = 1/2 spin sublevel of $^2\!E$ ground state of NV$^0$. 
            (a3) $K^i_{27}$: Electronic configuration of the $m_s$ = 0 spin sublevel of the $^3\!E$ NV$^-$ excited state ionizing to $m_s$ = 1/2 spin sublevel of the $^2\!A_2$ NV$^0$ state. 
            (a4) $K^i_{25}$: Electronic configuration of the $m_s$ = 0 spin sublevel of the $^3\!E$ NV$^-$ excited state ionizing to the quartet spin level of the $^4\!A_2$ NV$^0$ state. 
            (a5) Ionization via Auger recombination. An electron is promoted from the $a_{1}$ to the ${e_{x,y}}$ orbital by the abortion of a green photon. Subsequent absorption of an infrared photon promotes the electron from ${e_{x,y}}$ orbital to the conduction band. An Auger recombination process fills the hole in the $a_{1}$ orbital, releasing sufficient energy to detach an electron from the ${e_{x,y}}$, thereby leaving the system in the $^2\!E$ state.  
            (b) Available recombination processes, NV$^0\rightarrow$ NV$^-$. (b1) $K^\text{r}_{71}$: Electronic configuration of $^2\!A_2$ NV$^0$ excited state recombining to the $^3\!A_2$ ground state of  NV$^-$. 
            (b2) $K^\text{r}_{51}$: Electronic configuration of $^4\!A_2$ quartet state of NV$^0$ recombining to $^3\!A_2$ ground state of NV$^-$. (b3) $K^\text{r}_{73, 74}$: Electronic configuration of $^2\!A_2$ state of NV$^0$  and singlet states of  NV$^-$. The recombination pathways in (b3) are not included in our model (see text for details).    
            }
\end{figure*}

Insight into the ionization and recombination processes can be gleaned by considering the single-electron picture shown in Fig.\,\ref{fig:electronModel} for each of the available processes in Fig.\,\ref{fig:model}. In these diagrams, the effective energy of each state and the energy required to excite the electron (or hole) determine the total energy threshold for a given process. To identify processes that affect our measurements, we consider the energies of the 532\,nm (2.330\,eV), 966\,nm (1.283\,eV), and 1524\,nm (0.813\,eV) photons present in the experiments. 
We assume that the green laser initializes the system in the ${m_{\text{s}}=0}$ spin sub-level of NV$^-$ orbital ground state, and disregard spin-dependent ionization. 
We emphasize that processes such as photoionization from the NV$^-$ ground state $^3\!A_2$ to the NV$^0$ ground state $^2\!E$ and the NV$^-$ excited state $^3\!E$ to the NV$^0$ excited state $^2\!A_2$ are not feasible on the grounds that the energy threshold requirement exceeds the energy of the IR photons used in this study. These and other processes are discussed in detail below:
\begin{itemize}
    \item (a1 --- $K^i_{16}$) Photoionization from the $^3\!A_2$ ground state: 
    
    The electronic configuration of the $^3\!A_2$ state is $a_1^2e^2$. Therefore, after photoionization the system transitions into the ground state $^2\!E$ of NV$^0$ with electron configuration $a^2_1e^1$. The photoionization threshold IP($^3\!A_2\,\rightarrow\,^2\!E$) has been measured experimentally to be $2.65\,\text{eV}$\,\cite{Aslam2013,Bourgeois2017}. For our system, this process can only be accessed by multi-photon processes: 2-photon green excitation, 3-photon 966\,nm excitation, or 4-photon 1524\,nm excitation. Therefore, this photoionization process has been ignored. 

    \item (a2 --- $K^i_{46}$) Photoionization from the $^1\!E$ ground state:

    The photoionization threshold of this process is:
    \begin{equation}
    \begin{split}
        \text{IP}(^1\!E\,\rightarrow\,^2\!E) &= \text{IP}(^3\!A_2\,\rightarrow\,^2\!E) \\
        &- [E(^1\!E) - E(^3\!A_2)]\\
        &\approx 2.27\,\text{eV}\,.
    \end{split}
    \label{eq:1eE_to_2E}
    \end{equation}
    Here, we have used a theoretical estimate of [$E(^1\!E) - E(^3\!A_2$)] = 0.38\,eV\,\cite{Bhandari2021}. 
    This process has been omitted from our model for the following reasons. The energy threshold of $2.27$\,eV in Eq.\,\ref{eq:1eE_to_2E} can be satisfied by a single green photon, two 966\,nm photons, or three 1524\,nm photons. To assess the role of ionization from the $^1\!E$ state, we consider the following three scenarios, neglecting the three-photon process at 1524\,nm: (i) ionization induced by a single green photon, (ii) ionization induced by a single green photon together with two IR photons (966\,nm), and (iii) ionization from the $^1\!E$ state neglected entirely. We find that the resulting fits to the experimental data show no significant differences among these cases.
   One possible explanation for the negligible role of ionization from the $^1\!E$ state is the weak population trapping in this state. Under continuous-wave green illumination, the population predominantly cycles between the $m_s = 0$ sublevels of the ground and excited states. In the absence of resonant microwave driving, as in our experiment, the intersystem crossing (ISC) from the $^3\!E$ excited state to the singlet manifold remains low. As summarized in Table~\ref{table_rate}, the estimated ISC rate is much smaller than that of the competing ionization pathway ($K_\text{25,2-IR}^i$, discussed below). Consequently, the population in the singlet manifold is insufficient for the $^1\!E$ ionization channel ($K^i_{46}$) to play a significant role.
    Finally, photoionization from the NV$^-$ singlet excited state $^1\!A_1$ to the NV$^0$ ground state $^2\!E$ ($K^i_{36}$, not shown in Fig.\,\ref{fig:electronModel}) is neglected, on the account of the short lifetime and fast decay of the $^1\!A_1$\,\cite{Robledo2011,Tetienne2012,Qian2022PRA,Meirzada2018} relative to the $^1\!E$ state;  consequently, the population rapidly relaxes to $^1\!E$---it is therefore reasonable to treat the population as effectively residing in the $^1\!E$ state only.
   
    \item (a3 --- $K^i_{27}$) Photoionization from the $^3\!E$ excited state to $^2\!A_2$: 
    
    The electronic configuration of the $^3\!E$ state is $a^1_1e^3$. There are two processes available for removing one electron from either the $e_x$ or the $e_y$ level of NV$^-$, yielding two electron configurations with $a^1_1e^2$ as shown in (a3) and (a4) of Fig.\,\ref{fig:electronModel}, respectively.
    We first consider photoionization to the $^2\!A_2$ state---the photoionization threshold for this process takes the form:
    \begin{equation}
    \begin{split}
        \text{IP}(^3\!E\,\rightarrow\,{^2\!A_2}) &= \text{IP}(^3\!A_2\,\rightarrow\,^2\!E) - E^{\text{NV}^-}_\text{ZPL} \\
        &+ [E(^2\!A_2)-E(^2\!E)]\\
        &= 2.86\,\text{eV}\,.
        \end{split}
        \label{eq:3E_to_2A2}
    \end{equation}


    Here, $E^{\text{NV}^-}_\text{ZPL} = [E(^3\!E)-E(^3\!A_2)] = 1.946\,$eV and $E^{\text{NV}^0}_\text{ZPL} = [E(^2\!A_2)-E(^2\!E)] = 2.16\,$eV are the zero-phonon line energies of NV$^-$\,\cite{Doherty2011} and NV$^0$\,\cite{Doherty2013}, respectively, and IP($^3\!A_2\,\rightarrow\,^2\!E) = 2.65\,\text{eV}$\,\cite{Aslam2013,Bourgeois2017}.
    
    The photoionization process $^3\!E\,\rightarrow\,^2\!A_2$ in Eq. \ref{eq:3E_to_2A2} has a threshold energy 2.86\,eV that requires higher-order ($>$2) photon absorption processes and has therefore been ignored. 

    \item (a4 --- $K^i_{25}$) Photoionization from the $^3\!E$ excited state to $^4\!A_2$:
    
    The preferred and lowest energy state following photoionization from the $^3\!E$ excited state 
    is the quartet state $^4\!A_2$ of NV$^0$ with $a^1_1e^2$ electron configuration\,\cite{razinkovas2021photoionization,Siyushev2013,Felton2008}.
    The photoionization threshold for the process $^3\!E\,\rightarrow\,^4\!A_2$ is given by
    \begin{equation}
    \begin{split}
        \text{IP}(^3\!E\,\rightarrow\,^4\!A_2) &= \text{IP}(^3\!A_2\,\rightarrow\,^2\!E) - E^{\text{NV}^-}_\text{ZPL} \\
        &+ [E(^4\!A_2)-E(^2\!E)]\\
        &= 0.70\,\text{eV} + [E(^4\!A_2)-E(^2\!E)]\,.
        \end{split}
        \label{eq:3E_to_4A2}
    \end{equation}
    
    The value for $\Delta^0=$[$E$($^4\!A_2$)$-E$($^2\!E$)] (see Fig.\,3\,(b) main manuscript and Fig.\,\ref{fig:model}) is hitherto theoretically estimated to be 0.48--0.68\,eV\,\cite{razinkovas2021photoionization,Ranjbar2011}.
    Taking the lower bound of this range, the energy threshold for IP($^3\!E\,\rightarrow\,^4\!A_2)$ becomes $>$1.18\,eV (i.e. 0.70\,eV+0.48\,eV). 
    This threshold can be reached either by a single 966\,nm photon or by two 1524\,nm photons.
    Importantly, for the 966\,nm case, our analysis shows that both single-photon and two-photon ionization pathways from $^3\!E$ to $^4\!A_2$ must be included to accurately reproduce the experimental data. Specifically, single-photon ionization dominates at low IR  power, whereas two-photon ionization becomes significant at higher IR powers, and is crucial in reproducing the experiential data.
    We note that in the case of 1524\,nm, direct photoionization ($^3\!E\,\rightarrow\,^4\!A_2)$ is restricted to a two-photon process as IP($^3\!E\,\rightarrow\,^4\!A_2)> 1.18$\,eV exceeds the energy of a single 1524\,nm photon.
    Nevertheless, incorporating an effective single-photon ionization pathway is essential for reproducing the charge-state dynamics observed in the 1524\,nm dataset. A plausible physical origin of this contribution is discussed below in the context of ionization via Auger recombination\,\cite{Siyushev2013,Gali2019}  (see the next discussion, a5---$K^i_{26}$). 
    
    Going further, if the upper bound of $\Delta^0$ is taken to be 0.68\, eV, the resulting energy threshold for $\mathrm{IP}(^3\!E \rightarrow {}^4\!A_2)$ is 1.38\, eV. This threshold exceeds the energy of a single 966\, nm photon, thereby ruling out any single-photon ionization process, which our analysis requires. Consequently, consistency between the theoretical model and the experimental data constrains the upper bound of $\Delta^0$ to be $\leq$ 0.58\, eV and thus IP($^3\!E\,\rightarrow\,^4\!A_2)\leq\text{1.28}\,\text{eV}$.

    \item (a5 --- $K^i_{26}$) Ionization via Auger recombination:
    
    All of the the aforementioned ionization processes occur via direct photoionization. However, as shown in Refs.\,\cite{Siyushev2013,Guo2020APL,Huangfu2024}, ionization from $^3\!E$ to $^2\!E$ can occur as an Auger recombination process\,\cite{Gali2019}. Ionization via Auger recombination, which is illustrated in Fig.\,\ref{fig:electronModel}\,(a5), is a two-step process. Starting from the $^3\!A_2$ ground state, an electron is promoted from the $a_1$ orbital to the $e_{x,y}$ orbitals by absorbing a green photon. This process, which leaves a hole in the $a_1$, corresponds to the $^3\!A_2\rightarrow{^3\!E}$ transition. Next, absorption of a second photon, either green or IR, promotes an electron in the $e$ orbital to the conduction band. In the Auger recombination process, this electron in the conduction band falls back to occupy the hole in the $a_1$ orbital---the recombination energy is sufficiently large to detach and release an electron from the $e_{x,y}$\,\cite{Gali2019}. The resulting electron configuration is $a_1^2e^1$, corresponding to the $^2\!E$ state. Therefore, ionization via Auger recombination brings the population from the $^3\!E$ excited state of NV$^-$ to the ground state of NV$^0$.

    An upper limit on the energy separation between the $^3\!E$ state and the conduction band minimum can be estimated from IP($^3\!A_2\,\rightarrow\,^2\!E)-E^{\text{NV}^-}_{\text{ZPL}}\simeq0.7\,\text{eV}$\,\cite{Aslam2013,Bourgeois2017,Subedi2019}.  This energy difference can be met by either of the two IR wavelength considered here or by a single green photon. This absorption is followed by the Auger-recombination and subsequent removal of an electron from the $e_{x,y}$ orbital, thus leaving the NV center ionized.
    It is important to emphasize that from the $^3\!E$ excited state, direct photoionization ($^3\!E\rightarrow{^4\!A_2}$) and ionization via Auger recombination ($^3\!E\rightarrow{^2\!E}$) brings the system to different final states\,\cite{Gali2019,Wirtitsch2023}.  
\end{itemize}

We now turn to discuss recombination processes that bring NV$^0$ to NV$^-$. In the following discussion, we will consider excitation of holes to the valence band that is equivalent to capturing an electron from the valence band. The recombination processes require a hole in the molecular orbital ground state $a_1$ of NV$^0$ (Fig.\,\ref{fig:electronModel}\,(b)), to be filled without necessitating large energy threshold requirements. 
In this configuration, the population reside in the $^2\!A_2$ excited state or in the $^4\!A_2$ quartet state.
Recombination from the $^2\!E$ ground state of NV$^0$ requires a hole to be excited from the $e_x$ or $e_y$ molecular orbitals, which carries a larger energy threshold requirement and will therefore be ignored. We therefore only consider recombination processes to occur from the $^2\!A_2$ and $^4\!A_2$ states.

\begin{itemize}

\item 
(b1---$K^\text{r}_{71}$) Recombination from the $^2\!A_2$ excited state:

The electronic configuration of the $^2\!A_2$ state is $a_1^1 e^2$, consisting of one electron in the ground-state molecular orbital $a_1$ and two paired electrons occupying the degenerate excited-state orbitals ($e_x$ or $e_y$). The process in which a hole is excited from the $a_1$ level to the valence band---resulting in the configuration $a_1^2e^2$---is illustrated in (b1) of Fig.\,\ref{fig:electronModel}. Following recombination, the NV center relaxes to the $^3\!A_2$ ground state\,\cite{Wirtitsch2023, Thiering2024}.
The recombination threshold for this process can be estimated as
\begin{equation}
        \text{R}(^2\!A_2\rightarrow {^3\!A_2}) = \text{R}(^2\!A_2\rightarrow {^1\!E})- [E(^1\!E) - E(^3\!A_2)]\,.
\end{equation}

The recombination threshold $(^2\!A_2 \rightarrow {^1\!E})$ has not been adequately measured, with reports suggesting the threshold being less than 1.16 eV\,\cite{Ji2016}. By taking $[E(^1\!E) - E(^3\!A_2)] = 0.38\,$eV\,\cite{Bhandari2021}, the threshold R($^2\!A_2\rightarrow {^3\!A_2})$ becomes $<$0.78\,eV, which is $<\hbar\omega_\text{1524\,nm}$ (0.81\,eV).
In the discussion below (b2---$K^r_{51}$), we will further restrict the ranges of both R$(^2\!A_2 \rightarrow {^1\!E})$ and R($^2\!A_2\rightarrow {^3\!A_2})$.

The recombination pathway $^2\!A_2\rightarrow {^3\!A_2}$ is favored by Hund’s rule, which states that, for a given electron configuration, the state with the highest spin multiplicity ($2S + 1$, where $S$ is the total spin angular momentum) has the lowest energy. The $^3\!A_2$ state contains two unpaired electrons with parallel spins in the $e$ orbitals, yielding a spin multiplicity of 3. In contrast, the singlet states of NV$^-$, $^1\!A_1$ and $^1\!E$, have spin multiplicity 1, with antiparallel spins in the $e$ orbitals. Consequently, recombination from $^2\!A_2$ into these singlet states ($K^\text{r}_{73, 74}$) is energetically unfavorable and is therefore neglected in our model. For completeness, this recombination pathway is depicted in panel (b3) of Fig.\,\ref{fig:electronModel}.

We note that the recombination pathway $K^\text{r}_{72}$ from $^2\!A_2$ to the $^3E$ excited state of NV$^-$ is neglected, as it requires multiphoton ($>2$) processes at the wavelengths considered here and is therefore negligible. We will provide further support to this statement in the discussion (b2---$K^r_{51}$) below.

\item (b2---$K^r_{51}$) Recombination from $^4\!A_2$:

The electronic configuration of the $^4\!A_2$ state is $a^1_1e^2$. After recombination, the NV center transitions to the $^3\!A_2$ state. The recombination threshold for this process is described as follows:
    \begin{equation}
    \begin{split}
        \text{R}(^4\!A_2\,\rightarrow\,^3\!A_2) &= \text{R}(^2\!A_2\,\rightarrow\,^1\!E) - [E(^1\!E) - E(^3\!A_2)] \\
        &+ E^{\text{NV}^0}_\text{ZPL} - [E(^4\!A_2)-E(^2\!E)]\\
        &< \text{R}(^2\!A_2\,\rightarrow\,^1\!E)+ 1.30\,\text{eV}\,.
    \end{split}
    \label{eq:4A2_to_3A2}
    \end{equation}

Here, $E^{\text{NV}^0}_\text{ZPL} = [E(^2\!A_2)-E(^2\!E)] = 2.16\,$eV is the NV$^0$ zero-phonon line excitation energy\,\cite{Doherty2013}. From previous discussion on ionization, we have $\Delta^-=[E(^1\!E) - E(^3\!A_2)] = 0.38\,$eV\,\cite{Bhandari2021} (see the discussion on a2), and $\Delta^0=[E(^4\!A_2)-E(^2\!E)]>0.48\,\text{eV}$  (see a4), while from the ref.\,\cite{Ji2016}, we have $\text{R}({^2\!A_2}\rightarrow{^1\!E})<1.16\,\text{eV}$.
The close concordance between our fit and the data allows us to further limit this value.
From the model, we find that recombination from $^4\!A_2$ requires one green photon. 
In other words, $\text{R}(^4\!A_2\,\rightarrow\,^3\!A_2)\leq\hbar\omega_{532\,\text{nm}}$ (see Fig.\,\ref{fig:model} and Sec.\,\ref{sec:rate equations}). 
By using the equality and the lower bound $\Delta^0>0.48\,\text{eV}$ in Eq.\,\ref{eq:4A2_to_3A2}, we restrict $\text{R}({^2\!A_2}\,\rightarrow\,{^1\!E})<1.13\,\text{eV}$.

The lower limit for $\text{R}(^4\!A_2\,\rightarrow\,{^3\!A_2})$ can be estimated from 
\begin{equation}
    \text{R}(^4\!A_2\rightarrow{^3\!A_2}) \geq  \text{R}(^2\!A_2\rightarrow{^3\!A_2}) + E^{\text{NV}^0}_\text{ZPL} - \Delta^0\,.
    \label{eq:eq:4A2_to_3A2_second_eq}
\end{equation}
As before, using the lower bound $\Delta^0>0.48$\,eV yields $\text{R}(^4\!A_2\,\rightarrow\,{^3\!A_2})\geq\text{R}(^2\!A_2\rightarrow{^3\!A_2})+1.68\,\text{eV}$, which will always exceed the energy of two 1524\,nm photons ($2\hbar\omega_{1524\,\text{nm}}=1.63\,\text{eV}$). 
Therefore, since we limit the model to one- and two-photon processes, we restrict recombination from $^4\!A_2$ to $^3\!A_2$ to the absorption of one green photon or two 966\,nm photons only.

Finally, by using $\text{R}(^4\!A_2\,\rightarrow\,{^3\!A_2})=\hbar\omega_{532\,\text{nm}}$ and $0.48\,\text{eV}<\Delta^0<0.58\,\text{eV}$ in Eq.\,\ref{eq:eq:4A2_to_3A2_second_eq}, we find that $0.65\,\text{eV}<\text{R}(^2\!A_2\,\rightarrow\,{^3\!A_2})<0.75\,\text{eV}$, thus further restricting this recombination threshold.
Furthermore, as we alluded to above, the recombination pathway $K^\text{r}_{72}$ from $^2\!A_2$ to the $^3\!E$ excited state of NV$^-$ is given by 
\begin{equation}
    \text{R}(^2\!A_2\rightarrow{^3\!E})=\text{R}(^2\!A_2\rightarrow{^3\!A_2})+E^{\text{NV}^-}_{\text{ZPL}}\,.
    \label{eq:3A2to3E}
\end{equation}
From the constraints on $\text{R}(^2\!A_2\rightarrow{^3\!A_2})$, we find 
$2.60\,\text{eV}<\text{R}(^2\!A_2\rightarrow{^3\!E})<2.70\,\text{eV}$. This recombination threshold exceeds that of a single green and two IR photons for the wavelengths considered in this work, and this recombination pathway has therefore been ignored.
\end{itemize}

In Table\,\ref{tab:energies}, we summarize the energy values used to calculate the photoionization and recombination thresholds presented in this section. After selecting all the processes that meet the threshold criteria, we get two ionization process ($K^i_{25}$ and $K^i_{26}$) and two recombination processes ($K^r_{51}$ and $K^r_{71}$). Of these transitions,  $K^i_{25}$, $K^i_{26}$ and $K^r_{71}$ are accessible with a combination of IR and green excitation, whereas $K^r_{51}$ is accessible with green excitation only.

\renewcommand{\arraystretch}{1.2}
\begin{table}[t!]
\caption{Energy values used to calculate the photoionization and recombination thresholds in this section. Values denoted by $^*$ have been constrained in this work.}
\label{tab:energies}
\centering
\begin{minipage}{\columnwidth} 
\centering
\begin{tabular}{| c | c |}
\hline
\multicolumn{2}{|c|}{Photon Energy}\\
\hline
$\hbar\omega_{532\,\text{nm}}$ & $2.330\,\text{eV}$\\
$\hbar\omega_{966\,\text{nm}}$ & $1.283\,\text{eV}$\\
$\hbar\omega_{1524\,\text{nm}}$ & $0.813\,\text{eV}$\\
\hline
\hline
\multicolumn{2}{|c|}{Zero Phonon Line Energy}\\
\hline
$E^{\text{NV}^-}_\text{ZPL} = [E({^3\!A_2})-E(^3\!E)]$  & $1.946\,\text{eV}$\,\cite{Doherty2011}\\
$E^{\text{NV}^0}_\text{ZPL} = [E({^2\!A_2})-E(^2\!E)]$  & $2.16\,\text{eV}$\,\cite{Doherty2013}\\
\hline
\hline
\multicolumn{2}{|c|}{Energy Differences}\\
\hline
$\Delta^-=[E({^1\!E})-E(^3\!A_2)]$                              & $0.38\,\text{eV}$ (\cite{Bhandari2021})\\
$\Delta^0=[E({^4\!A_2})-E(^2\!E)]$                              & $0.48-0.58\,\text{eV}$ $^{*}$  (Eq. \ref{eq:3E_to_4A2})\footnote{\textcolor{black}{only the upper bound was restricted in this work, the lower bound is extracted from Ref.\,\cite{razinkovas2021photoionization,Ranjbar2011}.}}\\
\hline
\hline
\multicolumn{2}{|c|}{Photoionization Potential}\\
\hline
$\text{IP}({^3\!A_2}\rightarrow{^2\!E})$                        & $2.65\,\text{eV}$ (\cite{Aslam2013,Bourgeois2017})\\
$\text{IP}({^1\!E}\rightarrow\textcolor{black}{{^2\!E}})$  & $2.27\,\text{eV}$ (Eq.\,\ref{eq:1eE_to_2E})\\
$\text{IP}({^3\!E}\rightarrow{^2\!A_2})$                        & $2.86\,\text{eV}$ (Eq.\,\ref{eq:3E_to_2A2})\\
$\text{IP}({^3\!E}\rightarrow{^4\!A_2})$                        & $1.18-1.28\,\text{eV}$ $^*$ (Eq.\,\ref{eq:3E_to_4A2})\\
\hline
\hline
\multicolumn{2}{|c|}{Recombination Potential}\\
\hline
$\text{R}({^4\!A_2}\rightarrow{^3\!A_2})$                       & $\leq 2.33\,\text{eV}$ $^*$ (Eq.\,\ref{eq:4A2_to_3A2})\\
$\text{R}({^2\!A_2}\rightarrow{^1\!E})$                         & $<1.13\,\text{eV}$\ $^*$ (Eq.\,\ref{eq:4A2_to_3A2})\\
$\text{R}({^2\!A_2}\rightarrow{^3\!A_2})$                       & $0.65-0.75\,\text{eV}$ $^*$ (Eq.\,\ref{eq:eq:4A2_to_3A2_second_eq})\\
$\text{R}({^2\!A_2}\rightarrow{^3\!E})$                         & $2.60-2.70\,\text{eV}$ $^*$ (Eq.\,\ref{eq:3A2to3E})\\
\hline
\end{tabular}
\end{minipage}
\end{table}

\section{Seven-level rate equation model}
\label{sec:rate equations}
To quantitatively reproduce the data, we build a seven-level model that describes the NV$^-$--NV$^0$ system, as depicted in Fig.\,3(b) of the main manuscript. The following rate equations describe the coupling between these levels from the optical excitation, together with the requirement that the normalized population must be conserved:
\begin{equation}
\label{rateEq}
    \begin{split}
        \dot{p}_\text{1} &= -K_e^-p_1 + K_f^-p_2 + K_{41}p_4 + K_{71}^ip_5 + K^r_{51}p_5\,, \\
        \dot{p}_\text{2} &= - (K^i_{25} + K^i_{26}  + K_{23} + K_f^-)p_2 + K_e^-p_1 \\  
        \dot{p}_\text{3} &= -K_{34}p_3 + K_{23}p_2\,,\\
        \dot{p}_\text{4} &= -K_{41}p_4 + K_{34}p_3\,,\\
        \dot{p}_\text{5} &= -(K^r_{51}+K_{56})p_5 + K^i_{25}p_2 + K_{75}p_7\,,\\
        \dot{p}_\text{6} &= -K^0_ep_6 + K_{26}^ip_2 + K_{56}p_5 + K^0_fp_7\,,\\
        \dot{p}_\text{7} &= -(K^r_{71}+K^0_f+K_{75})p_7 + K^0_ep_6\,,\\
        \sum_{n=1}^7 p_n &= 1\,.\\
    \end{split}
\end{equation}
Here $p_n$ ($n \in [1,7]$) denotes the population of level with index $n$, and $K_{ij}$ represents the transition rate from level $i$ to $j$. The rates $K_f^{0,-}$  and $K_e^{0,-}$ are the NV$^{0,-}$ fluorescence decay and green laser power-dependent excitation rates, respectively. The ionization transition rates $K^i$ and the recombination transition rates $K^r$ are dependent on optical power and can be expressed using the following relations: 
\begin{equation}
\label{eq:coeff}
    \begin{split}
        K_{25}^i &=  \bar{K}_{25,1-966\,\text{nm}}^{i}(N_\text{IR})+ \bar{K}_\text{25,2-IR}^i (N_\text{IR}^2) +  \bar{K}_\text{25,1-G}^i (P_{\textrm{G}})\,,\\
        K_{26}^i &= \bar{K}_{26,1-\text{IR}}^i + \bar{K}_{26,1-\text{G}}^i \\
        K_{51}^r &=  \bar{K}_{51,2-966\,\text{nm}}^{r}(N_\text{IR}^2)+ \bar{K}_\text{51,1-G}^r (P_{\textrm{G}})\,,\\
        K_{71}^r &= \bar{K}_\text{71,1-IR}^r(N_\text{IR}) +  \bar{K}_\text{71,1-G}^r (P_{\textrm{G}})\,,
    \end{split}
\end{equation}

\begin{figure*}[tb!]
\begin{equation}
\label{eq:vector}
\begin{bmatrix}
-K_e^-  & K_f^-                                     & 0         & K_{41}    & K^r_{51}              & 0         & K^r_{71}\\
K^-_e   & -(K^i_{25}+K^i_{26}+K_{23}+K^-_f)   & 0         & 0         & 0                     & 0         & 0\\
0       & K_{23}                                    & -K_{34}   & 0         & 0                     & 0         & 0\\
0       & 0                                         & K_{34}    & -K_{41}   & 0                     & 0         & 0\\
0       & K^i_{25}                                  & 0         & 0         & -(K^r_{51}+K_{56})    & 0         & K_{75}\\
0       & K^i_{26}                           & 0         & 0         & K_{56}                & -K_e^0    & K_f^0\\
0       & 0                                         & 0         & 0         & 0                     & K_e^0     & -(K^r_{71}+K^0_f+K_{75})\\
1       & 1                                         & 1         & 1         & 1                     & 1         & 1
\end{bmatrix}
\begin{bmatrix}
p_\text{1} \\
p_\text{2} \\
p_\text{3} \\
p_\text{4} \\
p_\text{5} \\
p_\text{6} \\
p_\text{7} 
\end{bmatrix} =
\begin{bmatrix}
0 \\
0 \\
0 \\
0 \\
0 \\
0 \\
0 \\
1
\end{bmatrix}
\end{equation}
\end{figure*}

These rates are function of green power $P_{\textrm{G}}$ and the average number of IR photons circulating in the cavity, $N_\text{IR}$. The spatially averaged one and two photon IR excitation rates, $\bar{K}_{ij,p=1}$ and $\bar{K}_{ij,p=2}$, respectively, can be calculated using Eq.\ \ref{eq:K_single} and Eq.\ \ref{eq:K_multi}, respectively. The rates that depends on green excitation uses $I = P_\text{G}/A$ in Eq. \ref{eq:K1i}, where $P_\text{G}$ is the green laser power and $A$ is the area of the laser spot. For charge preserving internal rates, we utilize values from Refs.\,\cite{Tetienne2012,robledo2011spin,Meirzada2019} for NV excitation and fluorescence rates, and Refs.\,\cite{Tetienne2012, Qian2022PRA} for intersystem crossing rates. Note that each ionization or recombination rate is assigned a process-specific coefficient. We have also approximated the coefficients as constant parameters across the range of excitation power considered in this work.

\begin{table*}[!tb]
\centering
\caption{The rates used in this work. Internal NV transitions and green excitation rates are taken from literature. The rest of the rates are extracted from the fit to Eq.\,\ref{eq:vector}.}
 \label{table_rate} 
\begin{tabular*}{\textwidth}{|c@{\extracolsep{\fill}}l|cc||lrrl|}
\hline
\multicolumn{2}{|c|}{Fixed internal NV transitions} & \multicolumn{2}{c||}{\makecell{Fixed green excitation rates}} & \multicolumn{4}{c|}{Transition rates extracted from fits}\\
\hline
Transition & \makecell{Rate} & Transition & \makecell{Rate} & Transition & \makecell{Rate\\ (966\,nm)}& \makecell{Rate\\ (1524\,nm)} & Unit\\ 
 \hline
 \hline
 $K_f^-$    & 77\,MHz\,\cite{collins1983luminescence}       & $K_e^-$               & 10\,MHz\,/\,mW\,\cite{robledo2011spin,Meirzada2019}    & $\bar{K}_\text{25,1-G}^i$         & 35.11                 &  23.53                & kHz\,/\,mW \\  
 $K_f^0$    & 47.6\,MHz\cite{Beha2012}                      & $K_e^0$               & 18\,MHz\,/\,mW\,\cite{Qian2022PRA}    & $\bar{K}_\text{25,1-IR}^i$        & 488.23                &  $-$                  & Hz\,/\,photon \\
 $K_{23}$   & 7.9\,MHz\,\cite{Tetienne2012}                 & {}                    & {}                & $\bar{K}_\text{25,2-IR}^{i}$      & 2.04$\cdot 10^{-1}$   &  3.47$\cdot 10^{-5}$  & Hz\,/\,photon$^2$ \\
 $K_{34}$   & 1000\,MHz\,\cite{robledo2011spin}             & {}                    & {}                & $\bar{K}_\text{26,1-G}^i$         & 408.87                &  962.87               & Hz\,/\,mW \\
 $K_{41}$   & 5\,MHz\,\cite{robledo2011spin,Meirzada2018}   & {}                    & {}                & $\bar{K}_\text{26,1-IR}^i$        & 13.17                 & 1.75                  & Hz\,/\,photon \\
 {}         & {}                                            & {}                    & {}                & $\bar{K}_\text{51,1-G}^r$         & 744.00                & 677.49                & kHz\,/\,mW \\
 {}         & {}                                            & {}                    & {}                & $\bar{K}_\text{51,2-IR}^r$        & 5.64$\cdot 10^{-5}$   & $-$                   & Hz\,/\,photon$^2$\\
 {}         & {}                                            & {}                    & {}                & $\bar{K}_\text{71, 1-G}^{r}$      & 19.62                 & 8.19                  & kHz\,/\,mW \\
 {}         & {}                                            & {}                    & {}                & $\bar{K}_\text{71, 1-IR}^{r}$     & 1024.98               & 27.94                 & Hz\,/\,photon \\  
 {}         & {}                                            & {}                    & {}                & $K_{56}$                          & 363.34                & 363.68                & kHz \\  
 {}         & {}                                            & {}                    & {}                & $K_{75}$                          & 7.88                  &  12.13                & kHz \\  
 \hline
\end{tabular*}
\end{table*}

\begin{table*}[!tb]
\centering
\caption{Rates with uncertainty extracted by fitting the entire model. The rates are identical to the rates in Table.\,\ref{table_rate}. 
The uncertainty of the rates are obtained by computing the 95\% confidence interval on the entire model. The fit to the entire model yield an adjusted-$R^2$ value of 0.983 and 0.979 for 966\,nm and 1524\,nm, respectively.}
\label{table_rate_error} 
\begin{tabular*}{\textwidth}{|c@{\extracolsep{\fill}} | c l c| c l c |}
\hline
{} & \multicolumn{3}{c|}{966\,nm} & \multicolumn{3}{c|}{1524\,nm} \\
\hline
\ \ \ \ \ \ \ Transition\ \ \ \ \ \ \ \  & Rate & Unit & Uncertainty (\%) & Rate & Unit & Uncertainty (\%) \\ 
 \hline
 \hline
$\bar{K}_\text{25,1-G}^i$       & $35.11\pm 14.30$                          & kHz\,/\,mW            & $40.73$  & $23.53\pm7.41$                           & kHz\,/\,mW            & $31.50$ \\  
$\bar{K}_\text{25,1-IR}^i$      & $488.23\pm 192.64$                        & Hz\,/\,photon         & $39.46$  & $-$                                      & $-$                   & $-$ \\
$\bar{K}_\text{25,2-IR}^{i}$    & $\left(2.04\pm0.80\right)\cdot10^{-1}$    & Hz\,/\,photon$^2$     & $39.39$  & $\left(3.47\pm1.07\right)\cdot10^{-5}$   & Hz\,/\,photon$^2$     & $30.84$  \\
$\bar{K}_\text{26,1-G}^i$       & $408.87\pm 173.45$                        & Hz\,/\,mW             & $42.42$  & $962.87\pm336.20$                        & Hz\,/\,mW             & $34.92$  \\
$\bar{K}_\text{26,1-IR}^i$      & $13.17\pm 7.08$                           & Hz\,/\,photon         & $53.75$  & $1.75\pm0.63$                            & Hz\,/\,photon         & $36.00$  \\
$\bar{K}_\text{51,1-G}^r$       & $744.00\pm 305.68$                        & kHz\,/\,mW            & $41.09$  & $677.49\pm223.32$                        & kHz\,/\,mW            & $32.96$  \\
$\bar{K}_\text{51,2-IR}^r$      & $\left(5.64\pm2.32\right)\cdot10^{-5}$    & Hz\,/\,photon$^2$     & $41.09$  & $-$                                      & $-$                   & $-$       \\
$\bar{K}_\text{71, 1-G}^{r}$    & $19.62\pm 8.86$                           & kHz\,/\,mW            & $45.18$  & $8.19\pm2.78$                            & kHz\,/\,mW            & $34.00$  \\
$\bar{K}_\text{71, 1-IR}^{r}$   & $1024.98\pm 485.37$                       & Hz\,/\,photon         & $47.35$  & $27.94\pm9.79$                           & Hz\,/\,photon         & $35.04$  \\ 
$K_{56}$                        & $363.34\pm 198.26$                        & kHz                   & $62.93$  & $363.68\pm150.78$                        & kHz                   & $41.46$  \\
$K_{75}$                        & $7.88\pm 4.96$                            & kHz                   & $54.57$  & $12.13\pm5.03$                           & kHz                   & $41.49$  \\
\hline
\end{tabular*}
\end{table*}

\begin{table*}[!tb]
\centering
\caption{Rates extracted from single parameter fits. One parameter is free while the rest are fixed to the value in Table.\,\ref{table_rate}. The uncertainty and the $R^2$ value is computed from the 95\% confidence interval.}
\label{table_rate_error_single_param} 
\begin{tabular*}{\textwidth}{|c@{\extracolsep{\fill}} | c l c c| c l c c|}
\hline
{} & \multicolumn{4}{c|}{966\,nm} & \multicolumn{4}{c|}{1524\,nm} \\
\hline
Transition   & Rate & Unit & \makecell{Uncertainty \\ (\%)} & $R^2$  & Rate & Unit & \makecell{Uncertainty \\ (\%)} &  $R^2$ \\ 
 \hline
 \hline
$\bar{K}_\text{25,1-G}^i$       & $35.81\pm 0.14$                           & kHz\,/\,mW            & $0.39$    & $0.9833$      & $22.82\pm0.20$                            & kHz\,/\,mW        & $0.87$  & $0.9788$\\  
$\bar{K}_\text{25,1-IR}^i$      & $481.62\pm 3.29$                          & Hz\,/\,photon         & $0.68$    & $0.9831$      & $-$                                       & $-$               & $-$   & $-$\\
$\bar{K}_\text{25,2-IR}^{i}$    & $\left(1.97\pm0.02\right)\cdot10^{-1}$    & Hz\,/\,photon$^2$     & $0.96$    & $0.9832$      & $\left(3.30\pm0.042\right)\cdot10^{-5}$   & Hz\,/\,photon$^2$ & $1.27$  & $0.9788$\\
$\bar{K}_\text{26,1-G}^i$       & $291.75\pm 23.12$                         & Hz\,/\,mW             & $7.92$    & $0.9832$      & $1129.59\pm52.27$                         & Hz\,/\,mW         & $4.63$  & $0.9788$       \\
$\bar{K}_\text{26,1-IR}^i$      & $16.89\pm 0.66$                           & Hz\,/\,photon         & $3.92$    & $0.9833$      & $1.78\pm0.03$                             & Hz\,/\,photon     & $1.45$  & $0.9788$       \\ 
$\bar{K}_\text{51,1-G}^r$       & $744.00\pm 7.80$                          & kHz\,/\,mW            & $1.05$    & $0.9831$      & $719.81\pm8.12$                           & kHz\,/\,mW        & $1.13$  & $0.9789$      \\
$\bar{K}_\text{51,2-IR}^r$      & $\left(4.96\pm0.45\right)\cdot10^{-5}$    & Hz\,/\,photon$^2$     & $8.99$    & $0.9840$      & $-$                                       & $-$               & $-$   & $-$  \\
$\bar{K}_\text{71, 1-G}^{r}$    & $19.63\pm0.08$                            & kHz\,/\,mW            & $0.43$    & $0.9831$      & $7.86\pm0.09$                             & kHz\,/\,mW        & $1.12$  & $0.9789$       \\
$\bar{K}_\text{71, 1-IR}^{r}$   & $1032.68\pm 3.39$                         & Hz\,/\,photon         & $0.33$    & $0.9832$      & $28.08\pm0.18$                            & Hz\,/\,photon     & $0.64$  & $0.9788$       \\ 
$K_{56}$                        & $341.91\pm 6.52$                          & kHz                   & $1.91$    & $0.9832$      & $368.01\pm4.37$                           & kHz               & $1.19$  & $0.9788$       \\
$K_{75}$                        & $6.85\pm 0.12$                            & kHz                   & $1.69$    & $0.9835$      & $12.37\pm0.15$                            & kHz               & $1.23$  & $0.9788$      \\
\hline
\end{tabular*}
\end{table*}

To make predictions from the master equation, we define a population vector and calculate the steady state solution from Eq.\,\ref{eq:vector}. Treating the 966\,nm and 1524\,nm datasets independently, we perform a fit of this model to the data in Fig.\,3\,(a) in the main manuscript,  for varying green power.
The rates obtained from the fit, along with the fixed internal rates used in the model, are listed in Table\,\ref{table_rate}. 

We devise two methods to assess the quality of our fits and to extract the uncertainty in each fit parameter. 
First, we start by computing the 95\% confidence intervals of the entire model---we list the results in Table\,\ref{table_rate_error}.
In this method, all parameters are free to vary, with the exception of the fixed internal NV transition and green excitation rates as stated in Table\,\ref{table_rate}.
From the obtained fit results, we compute an adjusted-$R^2$ value of $0.983$ and $0.979$ for the 966\,nm and 1524\,nm datasets, respectively.
We emphasize that in the evaluation of these fits, we treat the two datasets (966\,nm and 1524\,nm) completely independently.
Therefore, the two fit results yield different values for the internal NV$^0$ rates ($K_{56}$ and $K_{75}$) and for the rates depending on green excitation only ($\bar{K}_\text{25,1-G}^i$, $\bar{K}_\text{26,1-G}^i$, $\bar{K}_\text{51,1-G}^r$ and $\bar{K}_\text{71, 1-G}^{r}$). However, all the extracted rate agrees to within the estimated uncertainty (see Table\,\ref{table_rate_error}).
As expected, the IR-assisted rates differs between the two IR wavelengths, which is explained by differences in the absorption cross-section (see discussion below).
We note that the measurements series for 966\,nm and 1524\,nm were performed on different days. Therefore, the difference in the green power dependent rates can be attributed to slight differences in focus and position of the green laser spot on the device, thus affecting the intracavity green power interacting with the NV centers. 
For future work, the uncertainty of the multivariate fit could be better constrained by performing additional measurements over a wider range of $P_\text{G}$.

As a second approach to assess the uncertainty, we next compute the 95\% confidence intervals of single parameter fits of the model.
In this method, the parameter of interest is allowed to vary, while all other parameters are fixed to the values in Table\,\ref{table_rate} obtained from the multivariate fit.
We list the resulting fit parameters, including the associated uncertainty and $R^2$ value in Table.\,\ref{table_rate_error_single_param}.
This process was utilized because of the cyclic nature of the processes in the model, which creates strong correlations between the rates, resulting in fit quality depending more strongly on the ratio of excitation and decay rates than on absolute rates.
As a result, when performing a multivariate fit, the uncertainty in some fitting parameters can be large (Table\,\ref{table_rate_error}).
As can be seen in Table\,\ref{table_rate_error_single_param}, this second method substantially reduces the fitting uncertainty.
The largest discrepancy in the extracted rates between the two datasets concerns the rate associated with $^4\!A_2$. The single parameter fit yield a factor of $\sim2$ difference in the estimation of $K_{75}$, where $K_{56}$ differs by $\sim8\%$.
However, it is important to acknowledge that this single parameter approach likely underestimates the uncertainty, as it does not easily capture the complex dynamics and interplay between the different rates.

\begin{figure}[tb!]
\includegraphics[width=\linewidth]{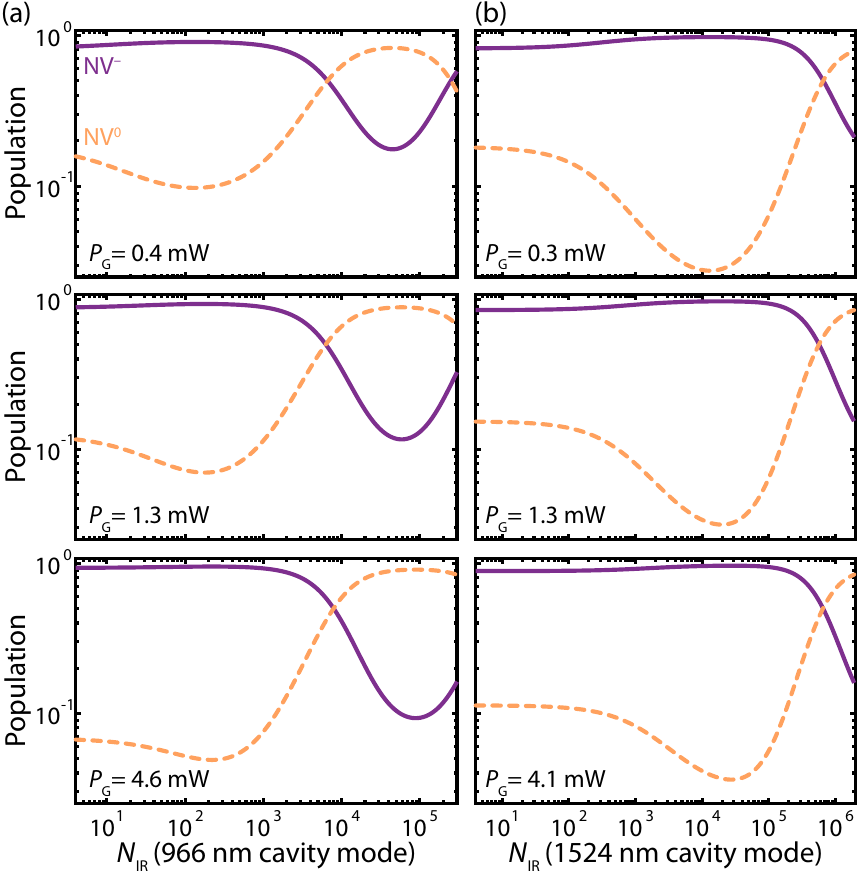}
		\caption{Prediction of the relative charge state population as a function of $N_{\text{IR}}$ for (a) 966\,nm and (b) 1524\,nm with increasing green power.
        For large $N_{\text{IR}}$ two-photon ionization dominates, and consequently trapping the population in the neutral charge state.}
        \label{fig:comp_Charge_State_population}
\end{figure}

The rate-equation model allows for making predictions on the charge state dynamics, and the relative population of the two charge states, as a function of $N_\text{IR}$ and green power (Fig.\,\ref{fig:comp_Charge_State_population}).
For $N_\text{IR}=0$, the relative charge state population is governed by the green power, and the population of NV$^-$ is found to increase with $P_{\text{G}}$. 
This is consistent with Fig.\,4\,(d) of the main manuscript. 
For 966\,nm, the increase in the population of NV$^-$ predicted for $N_\text{IR}\lesssim 10^2$ is governed by the ratio $\bar{K}_\text{71, 1-IR}^{r}\,/\,\left(\bar{K}_\text{25,1-IR}^i+\bar{K}_\text{26,1-IR}^i\right)$; recombination from the $^2\!A_2$ excited state of NV$^0$ is faster than single-photon ionization from the $^3\!E$ excited state of NV$^0$. 
A similar trend is observed for 1524\,nm, albeit for larger $N_\text{IR}\lesssim 10^4$.
Next, for large $N_\text{IR}$, the two-photon ionization process $^3\!E\rightarrow{^4\!A_2}$ with rate $\bar{K}_\text{25,2-IR}^{i}$ dominates. Consequently, the population is rapidly transferred to NV$^0$. 
For 966\,nm the population of NV$^-$ is predicted to increase for $N_\text{IR}\gtrsim10^5$. At this large intracity photon number, two-photon recombination from $^4\!A_2$ with rate $\bar{K}_{51,2-966\,\text{nm}}^{r}$ starts to depopulate the $^4\!A_2$ state, bringing population back to the $^3\!A_2$ ground state of NV$^-$.
Effectively, the population is now cycling from the $^3\!A_2$ ground state to the $^3\!E$ excited state of NV$^-$ by the absorption of a green photon, from where absorption of two-photon first pumps the population to the $^4\!A_2$ state, rapidly followed by pumping back to the $^3\!A_2$ ground state---this would be manifested experimentally as a complete quenching of PL.
Note that this two-photon recombination process is not energetically permitted by 1524\,nm photons.

Furthermore, the rate-equation model also provides new insight into the internal dynamics of the NV center.
First, from the $^3\!E$ excited state of NV$^-$, direct photoionization to the $^4\!A_2$ dominates over the Auger recombination process. 
This is apparent by the order-of-magnitude difference between the rates comprising $K_\text{25}^i$ compared to those comprising $K_\text{26}^i$, and holds for both 966\,nm and 1524\,nm.
Consequently, the uncertainty associated with the $K_\text{26}^i$ rates are comparatively large (see Tables\,\ref{table_rate_error} and \ref{table_rate_error_single_param}). 
Removing the $K_\text{26}^i$ from 966\,nm altogether does not significantly alter the fits, nor the extracted fit parameters. 
For 1524\,nm on the other hand, the single-photon process $\bar{K}_\text{26,1-IR}^i$ is crucial for reproducing the experimental data for low $N_{\text{IR}}$. 
For completeness, we have therefore included ionization by Auger recombination for both the IR wavelengths.

We next turn to examine the internal dynamics of NV$^0$ in more detail.
In the absence of any recombination processes, the total decay rate from the $^2\!A_2$ excited state of NV$^0$ is $K_\text{75}+K_\text{76}$. As the competition between these two rates will pose a limit on the optical cyclicity within NV$^0$, we next calculate the branching ratio from $^2\!A_2$. The radiative lifetime of NV$^0$ has been experimentally measured to be $\tau\sim21\,\text{ns}$\,\cite{Beha2012,Baier2020}, which yield $K_\text{76}=\frac{1}{\tau}=47.6\,\text{MHz}$. From this, we estimate the branching ratio from $^2\!A_2$ to the $^4\!A_2$ quartet state and to the $^2\!E$ ground state to be $\Gamma_{75}=\frac{K_\text{75}}{K_\text{75}+K_\text{76}}=0.017\,\%$ and $\Gamma_{76}=\frac{K_\text{76}}{K_\text{75}+K_\text{76}}=99.983\,\%$, respectively.
Here, we have used $K_\text{75}=7.88\,\text{kHz}$, as listed for 966\,nm in Table.\,\ref{table_rate}. 

\section{Absorption cross-sections}
The cross-sections, $\bar{\sigma}^{(p)}$, for one-and two-photon processes can be estimated from the Eq.\,\ref{eq:K_single}, and Eq.\,\ref{eq:K_multi}, respectively, by using the values extracted from the fit to the rate-equation model (Table\,\ref{table_rate}) alongside the key cavity parameters in Table\,\ref{table_simulation}.
The results are listed in Table.\,\ref{table_Cross_section}.
For these calculations, we have used the median of the simulated values for the confinement factors: $\Gamma_{966\,\text{nm}}=0.29$ and $\Gamma^{(2)}_{966\,\text{nm}}=0.14$ for the $966\,\text{nm}$ IR mode and $\Gamma_{1524\,\text{nm}}=0.38$ and $\Gamma^{(2)}_{1524\,\text{nm}}=0.20$ for the $1524\,\text{nm}$ mode.

\begin{table}[h!]
\caption{Absorption cross-section for the single-photon, $\bar{\sigma}$, and two-photon, $\bar{\sigma}^{(2)}$ transitions considered in this work. The cross-sections are calculated from Eq.\,\ref{eq:K_single} and Eq.\,\ref{eq:K_multi}, respectively, using the transition rates listed in Table.\,\ref{table_rate}.}
\label{table_Cross_section}

\begin{tabularx}{\columnwidth}{|c @{\extracolsep{\fill}} |c |c | l|}
\hline
{ } & \makecell{966\,nm} & \makecell{1524\,nm} & \makecell{Unit} \\ 
\hline
\hline
   \ \ \ \ \ \ $\bar{\sigma}_{25,\text{1-IR}}$\ \ \ \ \ \        & $2.55\times10^{-23}$     & $-$                   & $\text{m}^2$          \\
   \ \ \ \ \ \ $\bar{\sigma}_{25,\text{2-IR}}^{(2)}$\ \ \ \ \ \  & $3.37\times10^{-52}$     & $1.20\times10^{-55}$  & $\text{m}^4\text{s}$  \\
   \ \ \ \ \ \ $\bar{\sigma}_{26,\text{1-IR}}$\ \ \ \ \ \        & $6.88\times10^{-25}$     & $1.22\times10^{-25}$  & $\text{m}^2$          \\
   \ \ \ \ \ \ $\bar{\sigma}_{51,\text{2-IR}}$\ \ \ \ \ \        & $9.32\times10^{-56}$     & $-$               & $\text{m}^4\text{s}$          \\
   \ \ \ \ \ \ $\bar{\sigma}_{71,\text{1-IR}}$\ \ \ \ \ \        & $5.35\times10^{-23}$     & $1.95\times10^{-24}$  & $\text{m}^2$          \\ 
\hline
\end{tabularx}
\end{table}

\section{Time-resolved dynamics}
Finally, we measure the time-resolved response of NV$^-$ PL to a modulated 1524\,nm field. The 1524\,nm laser is passed through an electro-optic modulator with a finite extinction ratio of 25\,dB, before being input to the cavity resonance, allowing us to modulate the IR power in the cavity at frequency $\omega_{\text{EOM}}$. The output of the fiber taper was sent to a high-bandwidth photoreceiver connected to an oscilloscope, from which the time-dependent power in the cavity was monitored. During these measurements, NV$^-$ PL was measured using 4.1\,mW of green laser power input to the confocal microscope.  

\begin{figure}[tb!]
\includegraphics[width=\linewidth]{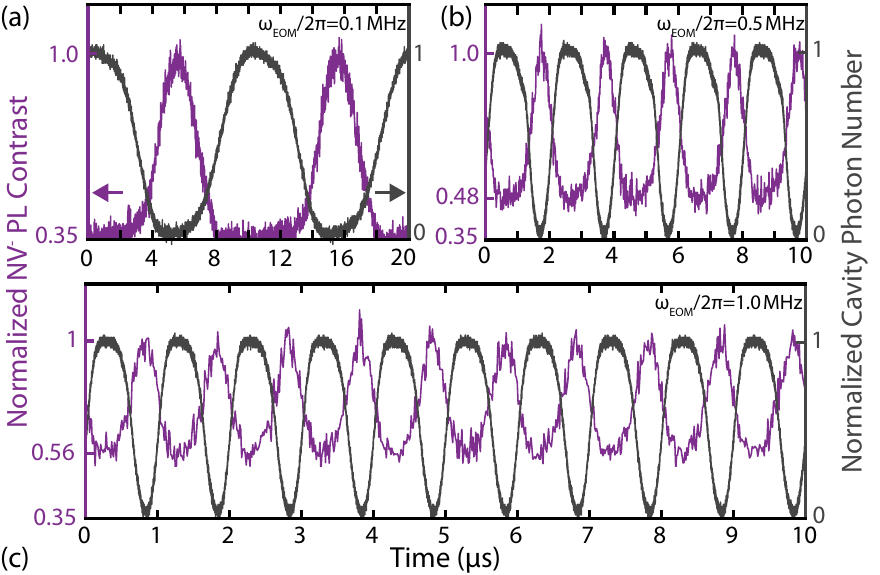}
		\caption{
		    \label{time}
            Time-resolved dynamics under modulation of the IR field. The blue lines show the normalized PL from NV$^-$ under 1524\,nm excitation. The black line shows the normalized interactivity photon number. The 1524\,nm laser was modulated at (a) $\omega_{\text{EOM}}/2\pi=0.1\,\text{MHz}$, (b) $0.5\,\text{MHz}$ and (c) $1.0\,\text{MHz}$. Note that the data in panel (a) is the same as displayed in Fig.\,4\,(e) of the main manuscript. 
	    }
\end{figure}

In Fig.\,\ref{time}, we show the time-varying IR power in the cavity and corresponding temporal variation in NV$^-$ PL for three different IR field modulation frequencies: $\omega_{\text{EOM}}/2\pi = 0.1~\text{MHz}$, 0.5 MHz, and 1 MHz. For modulation at $\omega_{\text{EOM}}/2\pi=0.1\,\text{MHz}$ (Fig.\,\ref{time}(a)), we observe a contrast in the NV$^-$ PL of $65\,\%$, which after considering the finite (25\,dB) extinction ratio of the EOM, is similar to the contrast observed in the measurement for constant IR power shown in Fig.\,\ref{figPL}(d), and in Fig.\,3(a) of the main manuscript. We next increase the modulation frequency to $\omega_{\text{EOM}}/2\pi=0.5\,\text{MHz}$ and $\omega_{\text{EOM}}/2\pi=1.0\,\text{MHz}$. The corresponding change in contrast is shown in Figs.\,\ref{time}(b) and (c), respectively, which show that as modulation frequency increases, the NV$^-$ PL contrast is reduced.
The decrease in the PL contrast with increasing modulation frequency is consistent with the total effective decay rate from the $^4\!A_2$ state of $2.1-4.2\,\text{MHz}$, which is given by sum of $K_{56}$ and $K_\text{51,1-G}^r$ (see Table\,\ref{table_rate}).

%